\documentstyle[aps,preprint]{revtex}

\newcommand{\thH}{\mbox{$\theta_H$}}
\newcommand{\thc}{\mbox{$\theta_c$}}
\newcommand{\thz}{\mbox{$\theta_0$}}
\newcommand{\epsn}{\mbox{$\epsilon$}}

\newcommand{\ben}{\begin{equation}}
\newcommand{\een}{\end{equation}}

\newcommand{\taubar}{\mbox{$\bar{\tau}$}}

\newcommand{\delbar}{\mbox{$\bar{\delta}$}}
\newcommand{\delcl}{\mbox{$\bar{\delta}_{\rm cl}$}}
\newcommand{\bebar}{\mbox{$\bar{\beta}$}}

\begin{document}
\draft
%\preprint{Submitted to Phys. Rev. B}
\title{Thermally Assisted Macroscopic
Quantum Tunneling of a
Ferromagnetic Particle in a Magnetic
Field at an Arbitrary Angle}

\author{Gwang-Hee Kim\cite{kim_address}}

\address{Department of Physics, Sejong University,
Seoul 143-747, KOREA}
\date{Received \hspace*{10pc}}
\maketitle

\thispagestyle{empty}

\begin{abstract}
At finite temperature
we study
the quantum tunneling of magnetization for a small
ferromagnetic particle with the biaxial symmetry
placed in a magnetic field at an arbitrary angle.
We present numerical
WKB exponent below the crossover temperature
in which the quantum tunneling is affected by the
thermal activation, and the
approximate form of the WKB exponent
around the crossover region. The effect of
quantum fluctuations on the thermal activation rate
beyond the crossover regime are discussed.\\
\end{abstract}

\section{Introduction}
\label{sec:int}

In the last decade, the problem of quantum tunneling of 
magnetization(QTM) in a single domain magnetic particle
has attracted a great deal of 
theoretical\cite{chu} and experimental\cite{her} interest.
There are two reasons for this.
One is the rapid development of the new field of nanomagnetism, 
in which a single domain magnetic particle can be prepared.
The other is that 
such a system meets the main criteria for observing 
macroscopic quantum tunneling (MQT). There are three conditions
for seeing MQT which can be stated qualitatively as follows.
First, the height and the width of the barrier should not be
large. Second, the effective moment of inertia associated with
the magnetization switching should not be too large. These
two conditions are required for a tunneling time
accessible to current experimental situations. Third, the crossover 
temperature at which the transition between
the thermal activation and the quantum
tunneling occurs should be of the order of
milliKelvin range or above.  Based on the criteria, it has been
believed that a magnetic field is a good external
parameter to make QTM observable. Also, tunneling dynamics
poses many interesting problems, particularly concerning
the role played by the thermal fluctuations at finite temperature.
This is the main subject of this paper, which deals with
thermal corrections to quantum tunneling rate of a
single domain ferromagnetic paricle in a magnetic field.

Consider a single domain ferromagnetic particle 
uniformly magnetized along
an easy axis determined by  
magnetocrystalline anisotropy. Since the anisotropy
generates two or more energetically 
equivalent orientations of the 
magnetization, at extremely low temperature
the direction of the magnetization $\bf{M}$ 
might change from one easy axis to the other easy axis by
the quantum tunneling process. 
However, according to the criteria, 
the rate of change of ${\bf M}$
is too small to be observed without controlling
the height and the width of the barrier as well as 
the effective moment of inertia intrinsically produced
by the magnetic anisotropy energy.
Applying an external magnetic field ${\bf H}$ in
a proper direction,
one of two equivalent orientations becomes metastable and
the magnitude of the physical quantities can be controlled 
in an appropriate way.
In this situation the change of ${\bf M}$ with time is
expected to be able to be observed and compared with
theroretical results. 
As temerature increases from zero,
thermal effects add to 
the quantum tunneling process.\cite{gar}
At sufficiently high temperature, 
the direction of the magnetization is changed by the pure
thermal activation whose rate is 
proportional to $\exp(-U/k_B T)$ where $U$ is an energy
barrier related to the magnetic anisotropy energy and the
external magnetic field. 
In the intermediate temperature region, it 
is expected that there
exists a crossover temperature $T_c$.
Around $T_c$
both thermal fluctuations and quantum tunneling 
coexist, namely, either the thermally 
activated quantum tunneling below $T_c$ or
the quantum mechanically assisted thermal 
activation process above $T_c$ can occur. 
Well below the
crossover temperature, i.e.,
at a temperature low enough to neglect the thermal 
activation process, the rate of change of ${\bf M}$
is soley due to quantum tunneling. 
\cite{mig,kim}

A number of theoretical
studies about QTM for the single domain ferromagnetic
particle
have been performed for several magnetocrystalline
anisotropy.\cite{chu}
The QTM problem with a magnetic field at an
arbitrary angle was firstly studied by Zaslavskii\cite{zas}
who calculated
the tunneling rate for the uniaxial symmetry by mapping
the spin system onto  a one-dimensional particle system.
For the same symmetry, Miguel and Chudnovsky\cite{mig}
have calculated
the tunneling rate within the imaginary time path 
integral method.\cite{lan} They have also
discussed the tunneling rate at finite temperature and suggested
experimental procedures. Kim and Hwang\cite{kim} have 
performed the
calculation based on instanton approach for biaxial 
and tetragonal symmetry. Their work presented the tunneling and 
oscillation rate between angles 
$\pi/2 \le \thH \le \pi$ at zero temperature, where
$\thH$ is the angle between the initial easy axis
and the external magnetic field. 
This paper
extends the tunneling rate for the biaxial symmetry
to a finite temperature. Within the instanton approach, we 
present the numerical results for the WKB exponent 
below the crossover temperature and their
approximate formulas around  
the crossover temperature. Also, we 
discuss the effect of
quantum fluctuations on the thermal activation rate
beyond the crossover regime.

This paper is structured in the following way. 
In Sec. \ref{sec:gen}, we introduce general formulation for
the tunneling rate based on the spin coherent state path
integral method. We discuss the angular dependence of the
critical field and the critical angle by using
approximate formula of the total energy in the small
$\epsn(=1-H/H_c)$ limit, where
$H_c$ is defined as a critical field at which
the barrier disappears. In Sec.
\ref{qtmzero}, we give a brief survey of the quantum tunneling
of magnetization in the presence of the magnetic field
with a general direction at zero temperature. In
Sec. \ref{qtmfin}, we consider thermal effects on the
quantum tunneling rates below $T_c$ and qunatum
corrections to thermal activation rates above 
$T_c$ and present their numerical results in the entire
temperature regime. 
In Sec. \ref{dis} we summarize the analytic results discussed and
provide actual estimates of such important quantities as the
tunneling rate, crossover temperature, and so on, for several
real materials.

\section{General formulation}
\label{sec:gen}

Consider the spin coherent state path integral
representation of the partition function given by
\ben
 Z(\beta \hbar)=
 \oint D [{\bf M}(\tau)]
 \exp(-S_E/\hbar),
\label{amp}
\een
where $\beta=1/k_B T$, the path sum is over all periodic paths
${\bf{M}}(\tau)={\bf{M}}(\tau+\beta \hbar)$, and
$S_E$ the Euclidean action\cite{lan} which includes
the Euclidean version of the magnetic Lagrangian $L_E$ as
\ben
 S_E[{\bf M}(\tau)]=\oint d\tau L_E[{\bf M}(\tau)].
\een
Due to strong exchange interaction in a single domain ferromagnetic
particle, the magnitude of the magnetization is a constant
$M_0$. For that reason the dynamical variable is the direction
of the magnetization, equivalently, $\theta(\tau)$ and
$\phi(\tau)$ in the spherical coordinates of $\bf{M}$.
Therefore, up to the normalization the functional measure in
Eq. (\ref{amp}) is equivalent to $D\Omega$ as
\begin{equation}
D \Omega =\lim_{\epsilon\rightarrow 0}
\prod_{k=1}^N\Bigl( {2J+1\over 4\pi}\Bigr)\sin{\theta_k}\,
d\theta_k\, d\phi_k\, ,
\label{mea}
\end{equation}
where $\epsn={\rm max}(\tau_{k+1}-\tau_k)$ and $J=M_0 V/\hbar \gamma$.
Here $\gamma=g \mu_B /\hbar$, with $g$ being the g-factor,
$\mu_B$ the Bohr magneton, $M_0$ the magnitude of the
magnetization and
$V$ the volume of the system. 
Since the tunneling rate $\Gamma$ of a metastable state
is proportional to the imaginary part of the free energy of the
system, Im$F$, and Im$Z \propto$ Im$F$, from Eq. (\ref{amp}) the 
tunneling rate 
in semiclassical limit, with an exponential accuracy, is 
\begin{equation}
 \Gamma
 \ \ \ \propto
 \ \ \ \exp(-S^{\rm min}_{E}(T)/\hbar),
\label{rate}
\end{equation}
where $S^{\rm min}_{E}(T)$ is obtained along the trajectory with
period $\beta \hbar$ that minimizes the Euclidean action 
\begin{equation}
  S_E=V \int_{0}^{\beta \hbar}\{i \frac{M_0}{\gamma}
  [1-\cos \theta(\tau)] \frac{d \phi(\tau)}{d \tau}
  + E[{\bf M}(\tau)]\} d \tau . \label{se}
\end{equation}
The first term in Eq. (\ref{se})
is the topological
Wess-Zumino term\cite{del,nie},
and the second term is the energy density, which 
is composed of the magnetic
anisotropy energy $E_a$ and the energy
given by an external magnetic field ${\bf H}$,
given by
\ben
 E[{\bf M}(\tau)]=E_a-{\bf M} \cdot {\bf H}.
 \label{tote}
\een
We have selected the biaxial symmetry
with magnetic anisotropy energy
$E_a=K_1 (\alpha_{1}^{2}+\alpha_{2}^{2})+K_2 \alpha_{2}^{2}$.
Here $\alpha_{i}$'s are the directional cosines and
$K_1>0$, $K_2>0$ are the parallel and transverse anisotropy constants,
respectively, whose relative magnitude can be of any value.
Applying the magnetic field in the $xz$ plane\cite{mig,kim}
and expressing
$\alpha_1$, $\alpha_2$ and $\alpha_3$ in 
spherical coordinates of ${\bf M}$,
the energy
(\ref{tote}) is expressed as
\begin{eqnarray}
 E[\theta,\phi]
  &=&K_1 \sin^2 \theta + K_2 \sin^2 \phi
  \sin^2 \theta  \nonumber \\ 
  & &  -H_x M_0 \sin \theta \cos \phi
  -H_z M_0 \cos \theta + E_0 ,
\label{eq:toten}
\end{eqnarray}
where $E_0$ is a constant to
make $E(\theta, \phi)$ zero at the initial
orientation. As will be seen later,
while there isn't an exact analog
of kinetic energy in the action (\ref{se}), there is an
effective moment of inertia in the dynamics of 
a single domain particle that is inversely
proportional to a linear combination of $K_2$
and $H_x$. 
For this reason we need
either the transverse anisotropy
constant $K_2$ or the magnetic field $H_x$
transverse to the initial easy axis
for quantum tunneling.
Otherwise, QTM is not observable due to the effective moment of
inertia whose magnitude becomes infinity.

Introducing the dimensionless
constants,
\begin{eqnarray}
 k_2=K_2/K_1,\ \ \ h_x=H_x/H_0, \ \ \
 h_z=H_z/H_0, \ \ \ H_0 = 2K_1/M_0,
 \label{eq:dimpar}
\end{eqnarray}
we obtain the energy (\ref{eq:toten})
written as
\begin{eqnarray}
 \bar{E}[\theta,\phi]
 &=&\frac{1}{2} \sin^2 \theta
 +{k_2 \over 2} \sin^2 \phi \sin^2 \theta \nonumber \\
 & & -h_x \sin \theta \cos \phi 
 - h_z \cos \theta + \bar{E}_0,
\label{eq:reden}
\end{eqnarray}
where $\bar{E}(\theta, \phi)=E(\theta, \phi)/2K_1$.
As noted in Eq. (\ref{eq:reden}), there are two
equivalent easy directions in the $xy$ plane, which are
$\phi=0$ and $\pi$ in the azimuthal angle
if $h_x=0$. 
We treat the problem of the easy directions
given by $\phi=0$ in the $xy$ plane.
Even though
we start the argument from $\phi=\pi$, the whole discussion is 
the same as the one from $\phi=0$ by replacing $\phi$
by $\pi+\phi$. So, the classical path becomes 
$\phi_{\rm cl}(\tau)=\phi_0(\tau)+n\pi$ where $n=0$ or 1
where $\phi_0(\tau)$ is the classical trajectory started from 
$\phi=0$.\cite{kim} However,
if the external field is not along 
the easy axis($z$ axis), i.e., $h_x \neq 0$, only
$\phi=0$ in Eq. (\ref{eq:reden}) is the easy direction in 
the $xy$ plane and $\phi=\pi$ not. 
Keeping this point in mind,
the total energy on the easy plane $\phi$ becomes
\begin{equation}
 \bar{E}(\theta,0)=\frac{1}{2} \sin^2 \theta
 -h \cos (\theta- \theta_H ) + \bar{E}_0,
 \label{en0}
\end{equation}
where we used $h_x=h \sin \theta_H$ and
$h_z=h \cos \theta_H$. 
Let us define $\theta_0$ to be the angle of metastable
state generated by the anisotropy energy and
the external magnetic field, and
$\theta_c$ the angle at which the barrier vanishes by
the external critical magnetic field $H_c$.
Then,
$\theta_0$ is determined by
$[d\bar{E}(\theta,0)/d\theta]_{\theta=\theta_0}=0$,
and $\theta_c$ and $h_c$ by both
$[d\bar{E}(\theta,0)/d\theta]_{\theta=\theta_c,h=h_c}=0$ and
$[d^2\bar{E}(\theta,0)/d\theta^2]_{\theta=\theta_c,h=h_c}=0$.
After simple calculations,
the dimensionless switching or critical field
$h_c(\equiv H_c/H_0)$ and the critical angle $\theta_c$
are expressed as\cite{mig,kim}
\begin{eqnarray}
 h_c &=&(\sin ^{2/3} \theta_H +
 |\cos \theta_H|^{2/3})^{-3/2}, \label{eq:hcb}\\
 \thc&=&{1\over2}\arcsin[{2|\cot\thH|^{1/3}\over
 1+|\cot\thH|^{2/3}}], \label{eq:sthcb}
\end{eqnarray}
where it should be noted that Eq. (\ref{eq:hcb}) is
the well-known Stoner-Wohlfarth expression.\cite{sto}
Simple analysis for Eq. (\ref{en0}) reveals that
the small value of $\epsn$ is favorable for the
magnetization switching because 
the height and the width of the barrier
are proportional to the power of $\epsn$. 
The situation can be achieved by
applying the external magnetic field close to
the critical field. Hence the problem will be considered in the 
small limit of $\epsn$, i.e., $\epsn \ll 1$.

Expanding $[d\bar{E}(\theta,0)/d\theta]_{\theta=\theta_0}=0$
about $\thc$, and 
using the relations
$[d\bar{E}(\theta,0)/d\theta]_{\theta=\theta_c,h=h_c}=0$
and
$[d^2\bar{E}(\theta,0)/d\theta^2]_{\theta=\theta_c,h=h_c}=0$,
we obtain the equation for $\eta(=\thc-\thz)$ as\cite{kim}
\begin{equation}
\sin(2 \theta_c) (\epsilon - \frac{3}{2} \eta^2)
- \eta \cos(2 \theta_c)(2 \epsilon - \eta^2)=0.
\label{eq:etaeq}
\end{equation}
Simple calculations show that $\eta$ is of the order of $\sqrt{\epsn}$.
Thus
the order of magnitude of the second term in
Eq. (\ref{eq:etaeq}) is smaller than that
of the first term by $\sqrt{\epsn}$ and the
value of
$\eta$ is determined by the first term
which leads to
$\eta \simeq \sqrt{2 \epsilon /3}$.
However, when $\thH$ is very close to
$\pi/2$ or $\pi$, $\sin (2 \theta_c)$
becomes close to zero,
and the second term is much larger than the
first term.
Therefore,
the value of $\eta$ is obtained from
the second term
when $\thH \simeq \pi/2$ or $\pi$,
which leads to
$\eta \simeq \sqrt{2 \epsilon}$
for $\theta_H \simeq \pi/2$ and $\eta \simeq 0$
for $\theta_H \simeq \pi$
from the detailed analysis of
the equations which
$\theta_0$ and $\theta_c$ satisfy.
Since the first term in (\ref{eq:etaeq})
is dominant in the range of value $\thc$ which
satisfies $\tan(2\thc) > O(\sqrt{\epsn})$,
$\eta \simeq \sqrt{2 \epsn /3}$
is valid for
$\pi/2 +O(\sqrt{\epsn}) < \thH < \pi -O(\sqrt{\epsn})$
from Eq. (\ref{eq:sthcb}).
Fig. \ref{etaval} shows that
this is checked by performing the
numerical calculation for the equations which
$\thz$ and $\thc$ satisfy.

Introducing a small variable $\delta (=
\theta - \theta_0)$, 
we obtain
an approximate form of
$\bar{E} (\theta, \phi)$ as
\ben
 \bar{E}(\delta, \phi)={k_2 \over 2} \sin^2 \phi \sin^2
 (\theta_0 + \delta) + h_x \sin (\theta_0+\delta)
 (1-\cos \phi)+\bar{E}_1(\delta),
\label{eq:totdel}
\een
where $\bar{E} (\theta, \phi)$ is represented as
$\bar{E}(\delta, \phi)$,
and  $\bar{E}_1(\delta)$ is a function of
only $\delta$ given by
\begin{eqnarray}
 \bar{E}_1 (\delta)&=&\frac{1}{4} \sin (2 \theta_c)(3 \delta^2
                 \eta - \delta^3) \nonumber \\
                  & & +\frac{1}{2} \cos
                 (2 \theta_c)[\delta^2 (\epsilon -
                 \frac{3}{2} \eta^2)+\delta^3 \eta -
                 \frac{\delta^4}{4}].
\label{eq:edelta}
\end{eqnarray}
Although $\cos(2\thc)$-term
in Eq. (\ref{eq:edelta}) looks smaller by a factor of
$\eta$ which is
of the order of $\sqrt{\epsilon}$, it can not be neglected
near $\theta_H = \pi/2$ and $\pi$ because $\sin ( 2 \theta_c)$
is almost zero for these regions of $\thH$.
In order to evaluate
the order of magnitude of each term in action
(\ref{se}) and simplify the calculations for
$\epsn \ll 1$, it is convenient to use
new scaled variables
\begin{eqnarray}
 \tilde{\tau} = \epsn^{\alpha/2} \  \omega_0 \tau, \ \ \
 \bar{\delta}=\delta/\sqrt{\epsn}, \ \ \
 \omega_0 = 2 \gamma K_1 /M_0.
 \label{eq:dimpar1}
\end{eqnarray}
Then, the Euclidean
action (\ref{se}) becomes
\begin{eqnarray}
  S_E[\delbar(\tilde{\tau}),
  \phi(\tilde{\tau})]&=&
  \hbar J \epsn^{-{\alpha \over 2}} 
  \int_{0}^{\tilde {\beta}} d \tilde{\tau}
   \{ i \epsn^{\alpha \over 2}
   [1-\cos (\theta_0 + \sqrt{\epsn}\bar{\delta})]
  \frac{d \phi}{d \tilde{\tau}} \nonumber \\
  & & +{k_2 \over 2} \sin^2 \phi \sin^2 (\thz+\sqrt{\epsn}\bar{\delta})
  +h_x \sin(\thz+\sqrt{\epsn}\bar{\delta})(1-\cos\phi)
  \nonumber \\
  & & + \frac{1}{4} \sin (2 \theta_c)\epsn^{3\over 2}(3 \delbar^2
  {\eta \over \sqrt{\epsn}} - \delbar^3)
  +\frac{1}{2} \cos (2 \theta_c) \epsn^2
  [\delbar^2 (1 - {3\eta^2 \over 2\epsn})
  +\delbar^3 {\eta \over \sqrt{\epsn}}
  -\frac{\delbar^4}{4}] \}.
\label{eq:newactn}
\end{eqnarray}
where 
$\tilde{\beta}=\epsn^{\alpha/ 2} \ \omega_0 \beta \hbar$ 
and the parameters $\alpha$ is 
fixed by the analysis of the
order of magnitude of each term in
the action (\ref{eq:newactn}) 
subject to the situation considered.

\section{QTM at zero temperature}
\label{qtmzero}

Consider the Euclidean 
action (\ref{eq:newactn}) at $T=0$\cite{kim}
by performing a simple dimensional analysis.
At $T=0$ ($\beta \rightarrow \infty$) the instanton
starts and ends at the metastable state $\delbar_{\rm cl}=0$,
$\phi_{\rm cl}=0$ for 
$\pi/2 + O(\sqrt{\epsn}) < \thH < \pi - O(\sqrt{\epsn})$,
in which $\bar{E}(\delbar_{\rm cl}=0, \phi_{\rm cl}=0)=0$. Thus,
from the energy conservation
the total energy $\bar{E}(\delta_{\rm cl},\phi_{\rm cl})$
in Eq. (\ref{eq:totdel}) becomes zero at any $\taubar$ for
$T=0$. Since 
$\sin(2 \thc)$-term ($\sim \epsn^{3/2}$) is 
larger than the last term ($\sim \epsn^{2}$)
by $\sqrt{\epsn}$ in Eq. (\ref{eq:newactn}),  
the first, second and third term should be of
the order of $\epsn^{3/2}$. 
Since $h_x$ is finite in the range of angle and
$\thz$ is not close to 0 and $\pi/2$, the small value of
$\phi$ in the third term of Eq. (\ref{eq:newactn}) contributes to
the path integral such that $\phi$ should be of the order of
$\epsn^{3/4}$. This
fact is valid for any value of $k_2$ because $h_x$ is not
zero in this range of angle and plays the role of the
effective transverse anisotropy component which is 
crucial for QTM. Therefore, from 
$\bar{E}(\delta_{\rm cl},\phi_{\rm cl})=0$,
the magnitude $\phi_{\rm cl}$ is approximately given by
\begin{equation}
 \phi^{2}_{\rm cl} = -\frac{2\bar{E}_1(\delta_{\rm cl})}{
         k_2 \sin^2(\theta_0+\delta_{\rm cl})+
         h_x \sin(\theta_0 + \delta_{\rm cl})}.
\label{eq:ephi}
\end{equation}
From Eq. (\ref{eq:edelta}) we get
$\bar{E}_1 (\delta) \sim \sin (2 \theta_c)(3 \delta^2
\eta - \delta^3)$ because $\cos (2\thc)$-term is much smaller
than $\sin (2\thc)$-term. 
Since $\eta \simeq \sqrt{2\epsn/3}$, and the critical angle
$\thc$ and the angle $\thz$ of metastable state
have values between 0 and $\pi/2$
in this range of angle, we get
$\phi_{\rm cl} \sim -i \sqrt{M \cot \thc}\epsn^{3/4}$,
where
the effective moment of inertia $M$ is\cite{comact}
\ben
 M={\sin \thc \over h_x+k_2 \sin \thc}.
\een
Thus, the approximate formula
of the classical Euclidean action is found to be
$S_{E}^{\rm cl} \sim i\hbar J \sin \thc \sqrt{\epsn}
\delcl \phi_{\rm cl} \sim
\hbar J \epsn^{5/4} \sqrt{M \sin \thc \cos \thc}$ 
from the dynamical part of the first term in 
Eq. (\ref{eq:newactn}), where $\delcl \sim O(1)$ and 
$\bar{E}(\delcl, \phi_{\rm cl})=0$.
By using the critical field (\ref{eq:hcb}) and the critical angle
(\ref{eq:sthcb}), the classical action is approximately 
\begin{equation}
S^{\rm cl}_{E} \approx \hbar J 
\epsilon^{5/4} g(\thH),
\label{eq:scl135}
\end{equation}
where 
\ben
g(\thH)=\frac{|\cot \theta_H|^{1/6}}
{\sqrt{1+\frac{K_2}{K_1}(1+|\cot \theta_H|^{2/3})}}.
\label{sthH}
\end{equation}

In the case of $\thH=\pi$, we have $\thc=\thz=\eta=0$ and 
$0 \leq \delta_{\rm cl} \leq 2\sqrt{\epsn}$ from
Eq. (\ref{eq:edelta}). In this situation, the action
(\ref{eq:newactn}) becomes
\begin{eqnarray}
  S_E[\delbar(\tilde{\tau}),
  \phi(\tilde{\tau})]&=&
  \hbar J \epsn^{-{\alpha \over 2}}
  \int^{\tilde{\beta}}_{0} d \tilde{\tau}
   \{ {i\over 2} \epsn^{\alpha+2 \over 2}
    \delbar^2 \
  (\frac{d \phi}{d \tilde{\tau}})
   +{k_2 \over 2} \epsn \delbar^2 \sin^2 \phi
  +{1 \over 2} \epsn^2 (\delbar^2 - {\delbar^4 \over 4}) \}.
\label{eq:actn180}
\end{eqnarray}
Since the last term in Eq. (\ref{eq:actn180}) is of the order 
of $\epsn^2$,
the order of magnitude of $k_2 \sin^2 \phi$ in the second term
should be $\epsn$. If $k_2$ becomes much smaller than $\epsn$,
in which the magnetic particle possesses 
the uniaxial symmetry,\cite{mig,zas} QTM
is not observable. This is understood from the fact as follows.
Consider the action (\ref{se}) as the dynamical system
in the Hamiltonian formulation which consists of the canonical
coordinate $\phi$ and $p_{\phi}=1-\cos\theta$.\cite{nie}
The Poisson bracket $\{p_{\phi},H \}$ which determines
the dynamics of the spin system with the Lagrangian (\ref{se})
becomes zero for $\thH=\pi$ because 
the Hamiltonian $H$ becomes a function 
of only $p_{\phi}$. Thus, 
$\theta$ does not depend on time.\cite{comalt}
If the value of $k_2$  is much greater than $\epsn$,
the small value of $\phi$,
i.e., $O(\phi) \sim \epsn^{1/2}$ mainly contributes to the
path integral, in which $\alpha$ is chosen to be 1. 
In this case,
a comparison of $k_2$-term with the last term
in Eq. (\ref{eq:actn180}) determines
$\phi_{\rm cl} \sim -i \sqrt{K_1 \epsn/K_2}$,
which leads to the least action 
$S^{\rm cl}_{E} \sim 
i\hbar J \epsn \delcl^2 \phi_{\rm cl} \sim
\hbar J \epsn^{3/2} \sqrt{K_1 / K_2}$.
The approximate forms of $S^{\rm cl}_{E}$ which we have discussed 
for $\pi/2 + O(\sqrt{\epsn}) < \thH < \pi - O(\sqrt{\epsn})$
and $\thH=\pi$ agree with the 
exact results in Ref. \cite{kim} up to the numerical factor
$N=16 \times 6^{1/4}/5$ and $8/3$,
respectively.\cite{comzero}
From above estimates we note that the dependence
of the WKB exponent on $\thH$, $\epsn$ and $K_2/K_1$
can be deduced from the simple
dimensional analysis based on the energy conservation
$\bar{E}(\delcl,\phi_{\rm cl})=0$ without solving
the equation of motion from $\delta S_{E}=0$.
However, in order to obtain the dependence of the WKB
exponent on $ K_1$, $K_2$, $\epsn$, and $\thH$ 
at finite temperature, we cannot make use of
the dimensional estimate as above, because
$\bar{E}(\delcl,\phi_{\rm cl}) \neq 0$. 
Thus, we need to calculate
the bounce solution by numerically solving
the equation of motion satisfied by the least trajectory.

\section{QTM at finite temperature}
\label{qtmfin}

Consider the Euclidean action
(\ref{eq:newactn}) with a periodic instanton for
$\pi/2 + O(\sqrt{\epsn}) < \thH < \pi - O(\sqrt{\epsn})$
and $\thH=\pi$.

\subsection{$\pi/2 + O(\protect\sqrt{\epsn}) 
< \thH < \pi - O(\protect\sqrt{\epsn})$}
\label{anglesub}

Since the $\sin (2 \thc)$-term  is much larger than
$\cos (2 \thc)$-term in Eq. (\ref{eq:newactn}),
the action is simplified as
\begin{eqnarray}
  S_E[\delbar(\tilde{\tau}),
  \phi(\tilde{\tau})]&=&
  \hbar J \epsn^{-{\alpha \over 2}} 
  \int^{\tilde{\beta}}_{0} d \tilde{\tau} 
   \{ -i \epsn^{\alpha+1 \over 2}
    \sin (\theta_0 + \sqrt{\epsn}\bar{\delta}) \ \phi \
  (\frac{d \delbar}{d \tilde{\tau}}) \nonumber \\
  & & +{k_2 \over 2} \sin^2 \phi \sin^2 (\thz+\sqrt{\epsn}\bar{\delta})
  +2 h_x \sin(\thz+\sqrt{\epsn}\bar{\delta})
  \sin^2({\phi \over 2})
  \nonumber \\
  & & + \frac{1}{4} \sin (2 \theta_c)\epsn^{3\over 2}(3 \delbar^2
  {\eta \over \sqrt{\epsn}} - \delbar^3) \},
\label{eq:actn135}
\end{eqnarray}
where $h_x \neq 0$ and $k_2$ has any value, and 
we performed the integration by part for the 
first term. 
Also, 
the total time derivative is neglected because
$\phi_{\rm cl}(-\tilde{\beta}/2)=\phi_{\rm cl}(\tilde{\beta}/2)=0$
by noting that from the Euler-Lagrange equation
$\phi_{\rm cl}$ is proportional to
$d \delcl /d \tilde{\tau}$ which is zero at
$\tilde{\tau}=\pm \tilde{\beta}/2$ due to the periodicity 
of $\delcl (\tilde{\tau})$.
As is well known for the crystal symmetry  such as
the form $K_z \alpha_{z}^{2}-K_y \alpha_{y}^{2}$,
the topological term
$d\phi/d\tau$ in Eq. (\ref{se}) plays important roles
in the presence of a clockwise or counterclockwise winding
over the barrier along the passage from $\phi=0$ to 
$\phi=\pi$ in macroscopic quantum coherence(MQC).\cite{del}
Since there is no such topological situation in our
quantum tunneling problem, the total derivative term
does not contribute to the classical action. 
In the action (\ref{eq:actn135}) the last term 
generates the potential barrier in
the problem. Thus, the magnitude of each term
should be the same order as that of the last term.
Noting that the order of magnitude of each term is
$O(\epsn^{(\alpha+1)/2} \phi)$, $O(\sin^2\phi)$,
$O(\sin^2(\phi/2))$ and $\epsn^{3/2}$, respectively,
the small values of $\phi$ contribute to
the path integral, 
and its order of magnitude is expected to be
$\epsn^{3/4}$, as was discussed at $T=0$. 
Therefore, an appropriate
choice of $\alpha$ is 1/2, which makes to each term to be
of the order of $\epsn^{3/2}$ in the integrand.

Performing the Gaussian integration over $\phi$ in the
partition function (\ref{amp})
and the measure (\ref{mea}), and introducing
the variable $\bar{\tau}=\tilde{\tau}\sqrt{\sin (2\thc)/M}$,
the remaining integral is of the form
\ben
 \int D[\delbar(\bar{\tau})]
  \exp(-S^{\rm eff}_{E}/\hbar),
\label{path180}
\een
where $\bar{\beta}=\tilde{\beta}\sqrt{\sin (2\thc)/M}=
\epsn^{1/4} \omega_0 \beta \hbar \sqrt{\sin (2\thc)/M}$
and the effective action is given by
\ben
  S^{\rm eff}_{E}[\delbar(\bar{\tau})]=
   \hbar J   \epsn^{5/4}  \sqrt{M \sin(2\thc)}
   \int_{0}^{\bar{\beta}} d \bar{\tau}
   [  {1\over 2}
   ({d \delbar \over d \bar{\tau}})^2+\bar{U}(\delbar)],
\label{actn135redu}
\een
where $\bar{U}(\delbar)=(\sqrt{6} \delbar^2 -\delbar^3)/4$.

The classical trajectory $\delbar_{\rm cl}(\taubar)$ with
period $\bebar$ which minimizes the effective action satisfies
\ben
{d^2 \delbar_{\rm cl} \over d \taubar^2} =
[{d \bar{U} \over d \delbar}]_{\bar{\delta}=\bar{\delta}_{\rm cl}}.
\label{clatra}
\een
and the corresponding total energy becomes
\ben
\bar{E}_{\rm tot}(\bebar)=
{1 \over 4} (\sqrt{6} \delbar^2 -\delbar^3)
-{1 \over 2}({d \delbar \over d \taubar})^2.
\label{toten-bounce}
\een
At zero temperature the classical trajectory 
becomes a regular bounce
$\delbar_{\rm cl}(\taubar)=
\sqrt{6}/\cosh^2[({3 \over 32})^{1/4}\taubar]$
with $\bar{E}_{\rm tot}(\bebar \rightarrow \infty)=0$. 
In order to include thermal corrections to quantum tunneling
rate, we need to consider the periodic solutions which satisfy
Eq. (\ref{clatra}). In fact, there exist two kinds of periodic
solution, $\delbar_m(=2\sqrt{6}/3)$ and the periodic motion 
$\delcl(\taubar)$ of the
particle in the inversed potential with total energy
$\bar{E}_{\rm tot}$, as is seen in Fig. 2.\cite{chu1}
For the constant solution $\delbar_m$, we obtain 
the classical action from Eq. (\ref{actn135redu})
\ben
S^{\rm cl}_{E}=S_0=\hbar J \epsn^{5/4} \sqrt{M \sin 2 \thc}
\bebar \bar{U}(\delbar_m)=\beta \hbar U_m,
\een 
where
\ben
U_m={8 \sqrt{6} \over 9}K_1 V \epsn^{3/2}
[{|\cot\thH|^{1/3}\over 1+|\cot\thH|^{2/3}}],
\een
and the corresponding escape rate 
\ben
\Gamma_0 \propto \exp(-S_0 /\hbar)=\exp(-U_m /k_B T),
\label{gamzero}
\een
which is the Boltzmann formula representing a pure
thermal activation. Let us define $T_c$ to be the temperature
at which the periodic solution $\delbar_{\rm cl}(\taubar)$
with period $\bebar$ approaches the limit $\delbar_m$.
Then, slightly below $T_c$, the thermal bounce 
$\delbar_{\rm cl}(\taubar)$ reduces to  small 
oscillations near
the bottom of the inversed potential $-\bar{U}(\delbar)$ 
shown in Fig. \ref{figsha}. 
In this case we take
only the first Fourier harmonics for the thermal bounce
because the next harmonics
are smaller near $T_c$,\cite{ive}
\ben
\delcl(\taubar)=\delbar_0(T)+\delbar_1(T) \cos (\bar{\omega}
\taubar),
\label{delbartc}
\een
with $\bar{\omega}=2 \pi/ \bebar$. 
The temperature dependence of $\delbar_0$ and $\delbar_1$
is crucial for both the magnitude of $T_c$ and
the dependence of $\delcl$ on $\taubar$. If $\delbar_1$ does
not depend on $T$, $\delcl$ becomes
a function of $\taubar$ through $\cos(\bar{\omega}_c \taubar)$
at $T=T_c$, which is contrary to the fact that $\delcl$ is a
constant $\delbar_m$ at $T_c$. Therefore, $\delbar_1$
should vanish at $T_c$ and have
a finite value below $T_c$. 
In order to obtain $T_c$ and 
the bounce solution for $T \lesssim T_c$, 
we substitute the harmonic type solution (\ref{delbartc}) into the
equation of motion for the thermal bounce, such as 
Eqs (\ref{inst135eq}) or (\ref{inst180eq}) subject to the situation
under consideration. 
We then obtain the temperature dependence of 
$\delbar_0$ and $\delbar_1$ by taking
the constant term
and the coefficient of $\cos(\bar{\omega}\taubar)$
to be zero,
and the relation 
$\bar{\omega}_c(=2 \pi/\bar{\beta}_c)
=-\bar{U}''(\delbar_m=2\sqrt{6}/3)$
which gives 
\ben
k_B T_c= {\hbar \omega_0 \over \pi} f(\epsn,\thH),
\label{harm}
\een
where
\ben
f(\epsn,\thH)=
({3 \over 8})^{1/4}
\epsn^{1/4}[{|\cot\thH|^{1/6}\over 1+|\cot\thH|^{2/3}}]
\sqrt{1+{K_2 \over K_1}(1+|\cot\thH|^{2/3})}.
\label{fth}
\een 
Since the thermal bounce (\ref{delbartc})
satisfies the equation of motion
\ben
{d^2 \delbar \over d \taubar^2} - {\sqrt{6} \over 2}
\delbar + {3 \over 4} \delbar^2 =0,
\label{inst135eq}
\een
the temperature dependence of $\delbar_0$ and $\delbar_1$ 
is given by
\ben
\delbar_0(T)={\sqrt{6} \over 3}[1+({T \over T_c})^2], \ \ \
\delbar_1(T)={2\sqrt{3} \over 3}\sqrt{1-({T \over T_c})^4},
\label{bouncetc}
\een
where at $T=T_c$
$\delbar(\taubar)=2\sqrt{6}/3(=\delbar_m)$ by noting that 
$\delbar_0 (T_c)=2\sqrt{6}/3$ and $\delbar_1(T_c)=0$.
Substituting in Eq. (\ref{actn135redu})
for the action, the approximate form of 
the minimal action can be written as
\ben
{S^{\rm min}_{E} \over \hbar} \approx {U_m \over k_B T} 
[1-3 ({Tc-T \over Tc})^2].
\label{actiontc}
\een
It is important that for $T \lesssim T_c$
the action $S^{\rm min}_{E}$
of the thermal bounce (\ref{delbartc}) is smaller than
the action $U_m /k_B T$ of the constant path $\delbar_m$.
Hence
below $T_c$, the functional integral for the decay is dominated by
the thermal bounce. 
This fact implies that $T_c$ is the crossover temperature
from the quantum-mechanical to the thermally activated decay. Noting 
that the Boltzmann formula (\ref{gamzero}) is derived from the
constant path $\delbar_{\rm cl}(\taubar)=\delbar_m$ and also
valid above $T_c$, the thermal bounce is degenerate into the
constant trajectory for $T \geq T_c$. 

In order to obtain the WKB exponent in the entire range of
temperatures less than $T_c$, 
we employ two equivalent approaches for the 
numerical calculation of the least action. First, we use
the equation of motion (\ref{inst135eq}) for the numerical
bounce solution, in which the action has been integrated over
$\phi$ with a periodic boundary condition 
$\delcl(\taubar=-\bebar/2)=\delcl(\taubar=\bebar/2)$ 
where $\bebar=2 \pi (2/3)^{1/4} (T_c/T)$.
As illustrated in Fig. \ref{figbou}, note that
numerical solutions become 
oscillatory formulas which are of the form (\ref{delbartc})
at temperatures slightly less than $T_c$. 
When inserted into the action 
(\ref{actn135redu}), 
the classical action can be obtained by performing the integration
numerically. Second, numerically solve
the coupled equation of motion of $\delcl$ and $\phi_{\rm cl}$ derived
from the Euler-Lagrange equation for 
the action (\ref{eq:actn135}) by incorporating the energy conservation
$\bar{E}(\delbar_{\rm cl}, \phi_{\rm cl})=0$. The second method
is reduced to the first one if $|\phi|$ is small. 
However, since the first method is not valid unless $|\phi|$ is 
small enough to perform the Gaussian integration over $\phi$, the 
second method is a more general approach to obtain the WKB exponent.
Although the first method is enough for the WKB exponent in our cases,
the second method is useful for checking the results of the
first. Thus, we used both approaches for the WKB exponent and
found that the results obtained from the first methods are 
identical to the one from the second method.

In Fig.\ref{figpha} we
present the phase diagram of $\delbar$ and
$\bar{\phi}(=\phi/\phi_0)$ where
$\phi_0= i g(\thH)\sqrt{2(1+|\cot \thH|^{2/3})} \epsn^{3/4}$
in the range of temperature
$0 \le T < T_c$. Note that
the orbit is oval for $0< T << T_c$ and circular near $T_c$ 
with a center at ($2\sqrt{6}/3,0)$.
Substituting $\delcl(\taubar)$ and $\phi_{\rm cl}$ into
the action (\ref{eq:actn135}), we obtain the WKB exponent 
which is shown in Fig. \ref{figtem}. 
Note that the approximate form (\ref{actiontc}) is valid at the
temperature close to $T_c$, and the WKB exponent is not
sensitive to $T$ for temperatures between
0 and $T_c$.
 
We now discuss the effect of quantum fluctuation
on the thermal activation at the temperature above $T_c$.
At temperature slightly above $T_c$ quantum
correction to the thermal activation becomes important. Its
effect is due to the extension of the regular
second-order action by terms up to the fourth order,
i.e., non-Gaussian terms in the action functional.
The quantum fluctuation, therefore, smears the transition in
the crossover region $|T_c-T| \ll T_c$, in which the
WKB exponent is of the form (\ref{actiontc}).
Even at higher temperature where thermal activation prevails,
the quantum effect is still incorporated into the preexponential
factor of the thermal activation rate in the form\cite{wei}
\ben
\Gamma={\omega_p \over 2 \pi} C_q \exp(-\beta U_m)
\een
where $\omega_p = \sqrt{\bar{U}''(\delbar=0)}$
is the frequency of oscillation around the
metastable minimum and $C_q$ a quantum mechanical
correction. The escape rate is enhanced by quantum 
fluctuations. This correction has been studied in great 
detail\cite{aff,wei}.  
Calculation of $C_p$ based on quantum fluctuation,
the decay rate which includes the 
quantum effect in this region has the form
\ben
\Gamma={\omega_p \over 2 \pi} \prod_{n=1}^{\infty}
{\omega^{2}_{n}+\omega^{2}_{p} \over \omega^{2}_{n}-\omega^{2}_{b}}
\exp(-\beta U_m),
\label{rateactq}
\een
where $\omega_n=2 \pi n /\beta \hbar$ and 
$\omega_b (= \sqrt{-\bar{U}''(\delbar_m)})$ is the barrier
frequency. This barrier frequency characterizes the width of the parabolic
top of the barrier. 
Using the identity
\ben
{1 \over 2 \sinh(\beta \hbar \omega/2)}
={1 \over \beta \hbar \omega} \prod_{n=1}^{\infty}
{\omega^{2}_{n} \over \omega^{2}+\omega^{2}_{n}},
\een
the decay rate (\ref{rateactq}) can be expressed as\cite{wei,gra}
\ben
\Gamma=({\omega_b \over 2\pi}) 
{\sinh(\beta \hbar \omega_p /2)
\over \sin(\beta \hbar \omega_b/2)}
\exp(-\beta U_m),
\label{thermalrate}
\een      
Since $\bar{U}''(0)=-\bar{U}''(\delbar_m)$ in this range of 
angles, we obtain $\omega_b=\omega_p=2 \omega_0 f(\epsn, \thH)$.
At temperatures slightly above the crossover region
$T-T_c \ll T_c$, we expand $\sin(\beta \hbar \omega_b/2)$
in Eq. (\ref{thermalrate})
about $T_c$. Noting that $\beta_c \hbar \omega_b =2 \pi$ from
the crossover temperature (\ref{harm}), we have 
$\sin(\hbar \omega_b /2 k_B T) \approx \pi (T-T_c)/T + 
\cdots$ and from  Eq. (\ref{thermalrate}) the escape rate is
given by
\ben
\Gamma \approx {\omega_0 \over \pi^2} f(\epsn,\thH)
{T_c \over T-T_c}
\sinh[\beta \hbar f(\epsn,\thH) \omega_0]
\exp(-\beta U_m).
\label{thermaltc}
\een

At temperatures well below the pure thermal activation regime,
but well beyond the crossover region $T-T_c \ll T_c$, 
we need to take the escape rate (\ref{thermalrate}) as
an exponential of a sum of logarithms and
expand each logarithm in power of $\beta \hbar$. Then, we obtain
the approximate form of the escape rate given by
\ben
\Gamma \approx {\omega_0 \over \pi} f^{2}(\epsn,\thH)
\exp \{-[{U_m\over k_B T}-{f^{2}(\epsn,\thH) \over 3}
({\hbar \omega_0 \over k_B T})^2]\}.
\label{thermalrate1}
\een     
As noted in Eq. (\ref{thermalrate1}),
the quantum effect which is inversely proportional
to $T^2$, is incorporated into the Boltzmann formula
($\beta U_m$) in the exponent which is indicative of 
quantum-mechanically assisted thermal activation process.

The expression (\ref{thermalrate}) which
includes the quantum fluctuation effects is valid in the
entire thermal activation regime beyond the crossover
region $T-T_c \ll T_c$. Its approximate forms
(\ref{thermaltc}) and (\ref{thermalrate1}) are valid
slightly and well above the crossover temperature, respectively. 
For a pure thermal activation regime 
in which $\beta$ becomes much small, we obtain the 
approximate relation
$\sinh(\beta \hbar \omega_p/2) \approx 
\sin (\beta \hbar \omega_b /2)$.
In this case 
the escape rate is given by
\ben
\Gamma = {\omega_0 f(\epsn, \thH) \over \pi} \exp(-\beta U_m),
\label{pthermal}
\een
which is just the Boltzmann formula 
for the pure thermal activation process from 
Eq. (\ref{thermalrate}).

Fig. \ref{figplot} illustrates that we superpose the 
approximate expressions for the tunneling rates on the 
numerical results and show the region of the validity of the
analytic results. Eq. (\ref{thermalrate1}) is applicable in 
a wide range of temperatures above $T_c$ but not appropriate as $T$
approaches the crossover region which includes $T_c$. 
Close to the crossover region, the
expression (\ref{thermaltc}) is valid in a narrow region. In the
crossover region Eq. (\ref{thermaltc})
is not appropriate because of singularity. In this case we need
to extend the second-order action by terms up to the higher
order in order to regularize the divergence, as previously noted, 
which leads to the WKB exponent to be of the form 
(\ref{actiontc}).\cite{comcross} When compared with the 
Boltzmann formula (\ref{pthermal}) based on the pure
thermal activation, quantum fluctuations increase as $T$
becomes close to $T_c$ from the pure thermal regime. Also,
note that while the tunneling rate at zero  temperature is 
valid in a wide range of temperature below $T_c$, 
there exist an appreciable amount of thermal fluctuations
for $T \lesssim T_c$.

\subsection{$\thH=\pi$}

When the external magnetic field is
opposite to the initial orientation,  
$h_x=\thz=\thc=0$ and the Euclidean action
becomes Eq. (\ref{eq:actn180}).
Estimating the order of magnitude of each term in the 
integrand of the action (\ref{eq:actn180}), 
$\alpha$ is chosen to be 1 and
$\phi \sim O(\epsn^{1/2})$ for the value of $k_2$ much larger 
than $\epsn$, as previously discussed at $T=0$.
Performing the Gaussian integration over $\phi$,
the effective action for the integral (\ref{path180}) is
given by
\ben
  S^{\rm eff}_{E}[\delbar(\bar{\tau})]=
   {\hbar J   \epsn^{3/2 } \over \sqrt{k_2}}
   \int_{0}^{\bar{\beta}} d \bar{\tau}
   [  {1\over 2}
   ({d \delbar \over d \bar{\tau}})^2
   + \frac{1}{2} (\delbar^2-{\delbar^4 \over 4})],
\label{actn180redu}
\een     
where we introduced $\taubar=\sqrt{k_2} \tilde{\tau}$
and $\bebar=\sqrt{k_2} \tilde{\beta}=
\epsn^{1/2} \omega_0 \beta \hbar \sqrt{K_2/K_1}$.
At $T=0$, the least trajectory for the action 
(\ref{actn180redu}) is $\delcl (\taubar)=
2/\cosh(\taubar)$ with $\bar{E}_{\rm tot}=\bar{U}(\delcl)
-{1 \over 2} ({d \bar{\delta}_{\rm cl} 
\over d \taubar})^2=0$ where
$\bar{U}(\delbar)=(4 \delbar^2-\delbar^4)/8$. 

As with sbusection \ref{anglesub}, 
the height of barrier $U_m=K_1 V \epsn^2$,
and the crossover temperature are
\ben
k_B T_c= {\hbar \omega_0 \over \pi} 
\sqrt{K_2 \over 2 K_1} \epsn^{1/2}.
\een
For $T$ slightly less than $T_c$, the thermal bounce 
is represented as
\ben
\delcl(\taubar)={\sqrt{10}\over 5} \sqrt{1+4({T\over T_c})^2}
+ {4 \over \sqrt{15}} \sqrt{1-({T\over T_c})^2}
\cos (\bar{\omega}\taubar),
\een
and its corresponding action 
\ben
{S^{\rm cl}_{E} \over \hbar} \approx {U_m \over k_B T}
[1-{32 \over 15} ({Tc-T \over Tc})^2].
\label{actiontcpi}
\een     

In the entire temperature range less than the crossover 
temperature, the WKB exponent is found 
by using the numerical solution for the equation given by
\ben
{d^2 \delbar \over d \taubar^2} - 
\delbar + {\delbar^3 \over 2} =0,
\label{inst180eq}
\een        
with a periodic boundary condition with a period 
$\bebar=(2\pi /\sqrt{2}) (T_c/T)$. 
Its result is similar to the one 
illustrated in Fig. \ref{figpha}. Using this solution, we
obtain the WKB exponent whose form is the same as Fig.
\ref{figtem}.

For the thermal activation regime beyond the crossover region
$T-T_c  \ll T_c$, it is
necessary to consider the escape rate 
(\ref{thermalrate}) by including 
quantum fluctuation with the frequency
$\omega_p=\epsn^{1/2}\omega_0 \sqrt{K_2/K_1}(=\omega_b/\sqrt{2})$. 
Then, the rate is 
approximately given by
\ben
\Gamma \approx {\sqrt{2} \omega_0 \over \pi^2}s(\epsn)
{T_c \over T-T_c}
\sinh[\beta \hbar s(\epsn) \omega_0]
\exp(-\beta U_m),
\een   
for the quantum correction regime
slightly beyond the crossover region $T-T_c \ll T_c$, and
\ben
\Gamma \approx {\omega_0 \over \pi} s(\epsn)
\exp \{-[{U_m\over k_B T}-{s^{2}(\epsn) \over 2}
({\hbar \omega_0 \over k_B T})^2]\},
\label{thermalrate2}
\een
for the temperature regime well beyond the crossover region,
where $s(\epsn)=\epsn^{1/2}\sqrt{K_2/2 K_1}$.
The shape of $\ln \Gamma(T)/\Gamma(0)$ is similar to
the results
shown in Fig. \ref{figplot} at all temperatures, 

\section{Discussion and conclusion}
\label{dis}

Table \ref{tabsum} summarizes the analytic results for the
range of angles discussed. Note that the $\epsn$-dependence of the
height of barrier, the crossover temperature and the WKB exponent
at $T=0$ are different for $\pi/2 + O(\sqrt{\epsn})< \thH < \pi - 
O(\sqrt{\epsn})$ and $\thH=\pi$.
The exponents of $\epsn$ in $U_m$, $T_c$ and $B(T=0)$
are equal to 1.5, 0.25 and 1.25 in a wide range of
angles between $\pi/2$ and $\pi$, and increase up to a
value of 2, 0.5 and 1.5 
if the applied field goes toward the opposite direction
of the initial magnetization ($\thH=\pi$), respectively.
The properties of the former are characteristic of the 
quadratic-plus-cubic potential whose form is well-known in 
Josephson systems, and those of the latter 
are characteristic of the 
quadratic-plus-quartic potential.\cite{leg}

The exponent of $\epsn$ in the WKB exponent is 
easily understood by 
noting that the 
WKB exponent is propotional to the height of barrier $U_m$ and
inversely propotional to the width of the barrier characterized by
$\omega_b$, i.e., $B \propto U/\hbar \omega_b$.
To be specific,
$B\propto \epsn^{3/2-1/4}$ for $\pi/2 + O(\sqrt{\epsn})
< \thH < \pi - O(\sqrt{\epsn})$ and
$B\propto \epsn^{2-1/2}$ for $\thH=\pi$. 
Denoting the WKB exponent $B(T=0)$ to be 
$c_B U_m /\hbar \omega_b$, $c_B$ is 36/5 for the range of
angle $\pi/2 + O(\sqrt{\epsn})
< \thH < \pi - O(\sqrt{\epsn})$ and 32/3 for $\thH=\pi$,
which are generally 
determined by the form of potential near liability.

The ratio of $T_c(\thH)$ to $T_c(135^{\circ})$ is
larger at $k_2=0$ than at finite $k_2$ for $\thH < 135^{\circ}$
and smaller for $\thH > 135^{\circ}$ (Fig. \ref{fig:tcbar}). The
position of the maximum of $\bar{T}_c$ moves from 
$\thH \rightarrow 101^{\circ}$ at $k_2 \rightarrow 0$
(uniaxial symmetry) to
$\thH \rightarrow 135^{\circ}$ at $k_2 \rightarrow \infty$,
and its magnitude $\bar{T}_c$ becomes smaller as 
$k_2$ increases. This behavior is easily understood from
Eq. (\ref{fth}) which gives
a maximaum value of the crossover temperature $T_c$ at
\ben
\theta^{T^{\rm{max}}_c}_H=\pi-\arctan[{K_2/K_1 \over
\sqrt{(K_2/K_1+0.5)^2+2}-1.5}]^{3/2},
\label{maxangle}
\een
as is illustrated in Fig. \ref{fig:tcmax}.

It is noted that the
WKB exponent $B$ at zero temperature can have the maximum value,
depending on the ratio of $K_2$ to $K_1$. For $K_1 \gg K_2$,
there is no maximal point in $B$ at zero temperature.\cite{mig,kim} 
Thus, there are some values of the ratio of the
anisotropy constants $K_2/K_1$, at which 
the WKB exponent changes slightly in a wide range of angles.
In the crossover region the quantum fluctuation is incorporated into
the WKB exponent as a parabolic form with a slightly
different curvature with respect to temperature in the range of angle
$\pi/2 + O(\sqrt{\epsn}) < \thH < \pi - O(\sqrt{\epsn})$
and $\thH=\pi$, which leads to the smearing of the 
WKB exponent, as is seen in Fig. \ref{figtem}. The quantum 
correction is also reflected in the thermal activation regime
with different coefficient of $(T_c/T)^2$ 
well above the crossover temperature
for $\pi/2 + O(\sqrt{\epsn})< \thH < \pi -
O(\sqrt{\epsn})$ and $\thH=\pi$.

To illustrate the above approximate results with concrete numbers 
we have collected in Table \ref{tabmag} and \ref{tabang} 
various values for several materials, the example with 
$K_1 =K_2$, $\rm{BaFe_{12-2x}Co_x Ti_x O_{19}(x=0.8)}$,\cite{wer1}
Tb,\cite{chu2} and Ni\cite{boz}. From Table \ref{tabang} it 
is evident that the typical magnitude of $\epsn$
for the magnetic system to tunnel out of
the barrier within resonable time is of the order of $10^{-3}$ or less,
and that the associated crossover temperature $T_c$ is
typically of the order of 100 mK or less. 
Noting that the inverse of the WKB
exponent $B^{-1}$ is the magnetic viscosity $S$ studied by
magnetic relaxation measurements,\cite{bar} the value
of $S(T=0)$ at the angle $\thH=135^{\circ}$ are of order
of 0.1-1 for the parameter $\epsn$ with the
magnitude $10^{-3}-10^{-4}$. Also,
the estimated values of $\theta_{H}^{T^{\rm{max}}_c}$ and 
$T^{\rm{max}}_c$ in Table \ref{tabang} 
can be compared with the experimental measurements
and their consistency will suggest the
possibility of QTM in these particles.

In conclusion, we have studied thermal effects on quantum tunneling
of magnetization placed in a magnetic field at an arbitrary angle.
We have presented the analytic forms of quantum tunneling rates at 
several temperature regimes and
performed the
numerical calculations in the entire temperature regime.
It is found that 
thermal corrections to  quantum tunneling rate is small
for $T \ll T_c$, but quite large for $T \sim T_c$.
This is because
the thermal activation process is strongly influenced by
the quantum fluctuation in this regime.
Furthermore, the
dependence of $T_c$ and magnetic viscosity on the parameters
$\thH$, $\epsn$ and $K_2/K_1$ 
is expected to be observed in future experiments.\\

\acknowledgments

I wishes to acknowledge the financial support of the Korea Research
Foundation made in the program year of 1997.

\pagebreak
\begin{table}
\caption{Summary of the results for the quantum tunneling 
         in the range of angle 
         $\pi/2 + O(\protect\sqrt{\epsn}) < \thH < 
         \pi - O(\protect\sqrt{\epsn})$
         and $\thH=\pi$. $B(=S^{\rm cl}_{E}/\hbar)$ is 
	 the WKB exponent
         at (a) zero temperature ($T=0$),\protect\cite{kim}
         (b) crossover region($T-T_c \ll T_c$), and
         (c) thermal activation region with
         quantum fluctuations ($T-T_c \gg T_c$).
         Here $t=T/T_c$,
         $m(\thH)=|\cot\thH|^{1/3}/ (1+|\cot\thH|^{2/3})$,
         $g(\thH)$ is given by Eq. (\protect\ref{sthH}) and 
         $f(\epsn,\thH)$ by Eq. (\protect\ref{fth}).}
\label{tabsum} 
\end{table} 
\begin{center} 
\begin{tabular}{|c|c|c|}  \hline 
$ $ & $\pi/2 + O(\sqrt{\epsn}) < \thH 
< \pi - O(\sqrt{\epsn})$ & 
$\thH=\pi$   \\ \hline 
Form of potential&
${\sqrt{6} \over 4} \delbar^2-\delbar^3$ &
${1\over 2} \delbar^2-{1\over 8}\delbar^4$ \\ 
near metastable point& & \\
$\alpha$   & 1/2 & 1 \\ 
$U_m/K_1 V$ & 
${8 \sqrt{6} \over 9}  \epsn^{3/2}m(\thH)$ & 
$\epsn^2$ \\ 
$\omega_b/\omega_0$ & $2 f(\epsn, \thH)$ &
$\epsn^{1/2}\sqrt{2 K_2/K_1}$ \\
$k_B T_c/(\hbar \omega_0/\pi)$ & 
$f(\epsn, \thH)$ & $\sqrt{K_2 \over 2 K_1} \epsn^{1/2}$ \\ 
$\delbar_0(T)$ & ${\sqrt{6} \over 3}(1+t^2)$ & 
${\sqrt{10} \over 5}(1+4 t^2)^{1/2}$ \\ 
$\delbar_1(T)$ & ${2\sqrt{3} \over 3} (1-t^4)^{1/2}$ & 
${4 \over \sqrt{15}} (1-t^2)^{1/2}$ \\ 
(a) $B$ & 
${16 \times 6^{1/4} \over 5}J\epsn^{5/4}g(\thH)$ & 
${8 \over 3} J \epsn^{3/2}\sqrt{K_1/K_2}$ \\ 
(b) $B$ & ${U_m \over k_B T}[1-3(1-t)^2]$ & 
${U_m \over k_B T}[1-{32 \over 15}(1-t)^2]$ \\ 
(c) $B$ & ${U_m \over k_B T}-{\pi^2 \over 3 t^2}$ &
${U_m \over k_B T}-{\pi^2 \over 4 t^2}$ \\ 
\hline
\end{tabular} 
\end{center}

\pagebreak

\begin{table}
\caption{Magnetization $M_0$, easy-axis anisotropy constant $K_1$,
hard-axis anisotropy constant $K_2$, critical magnetic field
$H_c$, and characteristic frequency $\omega_0$ for various
materials, where in case of Ni the shape-induced contribution
to the hard axis anisotropy is obtained for the slab geometry.}
\label{tabmag}
\end{table}
\begin{center}
\begin{tabular}{|c|c|c|c|c|c|c|c|}  \hline
  & $M_0$ & $K_1$ & $K_2$ & $V$ & $H_c(135^{\circ})$  &
$H_c(180^{\circ})$ & $\omega_0$ \\
  & [$\rm{emu/cm^3}$] & $10^5 [\rm{erg/cm^3}$] &
$10^5 [\rm{erg/cm^3}$] & [$\rm{10^3 nm^3}$] & [$\rm{10^{3}Oe}$] &
[$\rm{10^3 Oe}$] & [$\rm{10^{11}/sec}$] \\  \hline
$K_1=K_2$ & 500 & 100 & 100 & 1 & 20 & 40 & 7.04 \\
$\rm{BaFe_{12-2x}Co_x Ti_x}$ & 340 &
12 & $\sim 0$ & 3.14 & 3.53 & 7.06 & 1.24 \\
$\rm{O_{19}(x=0.8)}$\cite{wer1} & & & & & & &  \\
Tb\cite{chu2} & $\sim 100$ & $\sim 10$ & $\sim 10^3$ & 
$\sim 1$ & 10 & 20 & 3.54 \\
Ni\cite{boz} & 508 & 8 & 16 & $\sim 1$ & 1.575 & 3.15 & 0.557 \\
\hline
\end{tabular}
\end{center}

\pagebreak

\begin{table}
\caption{Angle for the maximal WKB exponent $\theta^{B_{\rm{max}}}_H$ and
the maximal crossover temperature $\theta^{T^{\rm{max}}_c}_H$, magnetic
field parameter $\epsn$, the height of barrier $U_m$ at 
$\thH=135^{\circ}$ and $\thH=180^{\circ}$, the WKB exponent $B(T=0)$ at
$\thH=\theta^{B_{\rm{max}}}_H$ and $135^{\circ}$, the crossover
temperature $T_c$ at $\thH=\theta^{T^{\rm{max} }_c}_H$, and 
inverse tunneling rate $\Gamma^{-1}(T=0)$ at $\thH=135^{\circ}$
for various materials.}
\label{tabang}
\end{table}

\begin{center}
\begin{tabular}{|c|c|c|c|c|c|c|c|c|c|}  \hline
  & $\theta^{B_{\rm{max}}}_H$ & $\theta^{T^{\rm{max}}_c}_H$ &
$\epsn$ & $U_m(135^{\circ})$  &
$U_m(180^{\circ})$ & $B_{\rm {max}}$ & 
$B(135^{\circ})$ & $T^{\rm {max}}_c$ & 
$\Gamma^{-1}(0,135^{\circ})$ \\ 
& [${}^{\circ}$] & [${}^{\circ}$]& &[K]&[K]& & & [mK] & [sec] \\
\hline
$K_1=K_2$ & 161 & 113 & $10^{-1}$ & $2.50\times 10^3$ & 725&
$4.50 \times 10^{3}$ & $4.34 \times 10^3$ & 669 & 
$6.30\times 10^{1870}$ \\
 &  &  & $10^{-2}$ & 78.9 & 7.25 & 253 & 244 & 376 & 
$6.38\times 10^{92}$ \\
 &  &  & $10^{-3}$ & 2.50 & $7.25\times 10^{-2}$ & 14.2 & 
13.7 & 211 & $4.59\times 10^{-7}$ \\
 &  &  & $10^{-4}$ & $7.89 \times 10^{-2}$ & $7.25\times 10^{-4}$ & 
0.799 & 0.772 & 119 & $8.35\times 10^{-12}$ \\
\hline
$\rm{BaFe_{12-2x}}$ &
180 & 101 & $10^{-1}$ & 940 & 273 &
$\infty$ & $1.62 \times 10^4$ & 76.9 & $1.73\times 10^{7022}$ \\
$\rm{Co_x Ti_x O_{19}}$ &
 &  & $10^{-2}$ & 29.7 & 2.73 & $\infty$ & 909 & 43.2 & 
$2.08\times 10^{382}$ \\
$\rm{(x=0.8)}$ & 
 &  & $10^{-3}$ & 0.94 & $2.73\times 10^{-2}$ & $\infty$ &
51.1 & 24.3 & $4.09\times 10^{10}$ \\
 & &  & $10^{-4}$ & $2.97\times 10^{-2}$ & $2.73\times 10^{-4}$ & 
$\infty$ & 2.88 & 13.7 & $3.50\times 10^{-10}$ \\
\hline
Tb & 135 & 135 & $10^{-1}$ & 250 & 72.5 &
107 & 107 & 2.67 K & $4.18\times 10^{32}$ \\
 &  &  & $10^{-2}$ & 7.89 & 0.725 & 6.00 & 6.00 & 1.50 K & 
$4.29\times 10^{-11}$\\
 &  &  & $10^{-3}$ & 0.25 & $7.25\times 10^{-3}$ & 0.337 &
0.337 & 845 & $1.12\times 10^{-12}$ \\
 &  &  & $10^{-4}$ & $7.89\times 10^{-3}$ & $7.25\times 10^{-5}$ &
$1.90\times 10^{-2}$ & $1.90\times 10^{-2}$ & 475 & 
$6.09\times 10^{-12}$ \\
\hline
Ni & 151 & 120 & $10^{-1}$ & 200 & 58.0 &
$3.48\times 10^3$ & $3.44\times10^3$ & 67.0 & 
$9.46\times 10^{1480}$ \\
 &  &  & $10^{-2}$ & 6.31 & 0.580 & 195 & 193 & 37.7 & 
$4.98\times 10^{71}$\\
 &  &  & $10^{-3}$ & 0.2 & $5.80\times 10^{-3}$ & 11.0 & 10.9&
21.2 & $3.06\times 10^{-7}$ \\
 &  &  & $10^{-4}$ & $6.31\times 10^{-3}$ & $5.80\times 10^{-5}$ &
0.617 & 0.612 & 11.9 & $7.83\times 10^{-11}$ \\
\hline
\end{tabular}
\end{center}

\pagebreak

\begin{figure}
 \caption{$\eta(=\thc-\thz)$ as a function
	  of $\thH$ for $\epsn=$ (a) 0.01 and
	  (b) 0.001. Note that 
	  $\eta \approx \protect\sqrt{2\epsn/3}$ is valid
	  in a wide range of angles, i.e., 
          $\pi/2 + O(\protect\sqrt{\epsn}) < \thH <
          \pi - O(\protect\sqrt{\epsn})$.}
 \label{etaval}
\end{figure}

\begin{figure}
 \caption{(a) The shape of the potential $\bar{U}(\delbar)$ and
          (b) the shape of the inverted potential $-\bar{U}(\delbar)$
          for the range of angle 
          $\pi/2 + O(\protect\sqrt{\epsn}) < \thH < 
          \pi - O(\protect\sqrt{\epsn})$. In (a) the height of 
          barrier $\bar{U}(\delbar_m =2\protect\sqrt{6}/3)$ 
          is $2\protect\sqrt{6}/9$ at $\bar{\delta}_m
          =2\protect\sqrt{6}/3$.}
 \label{figsha}
\end{figure}

\begin{figure}
 \caption{The bounce solution obtained by numerically solving Eq. 
         (\protect\ref{inst135eq}) for several temperatures 
         $t=$ (a) 0.95, (b) 0.98, (c) 0.99
         and (d) 0.999, where $t=T/T_c$. 
         The dashed curves in each temperature are drawn
         from the oscillatory solution 
         (\protect\ref{delbartc}) for comparison.
         Notice that as $T$ becomes close to $T_c$,
         the approximate formulas from Eq. (\protect\ref{delbartc}) are
         compatible with numerical solutions.}
 \label{figbou}
\end{figure}

\begin{figure}
 \caption{The phase diagram $\bar{\phi}$
          vs. $\delbar$ at 
          $t=$ (a) 0, (b) 0.82, (c) 0.97,
          and (d) 0.999. Note that 
          as $T$ increases from zero,
          the orbit is changed
          from an egg-shape to a circle.}  
 \label{figpha}
\end{figure}

\begin{figure}
 \caption{The temperature dependence of the magnetic viscosity,
          $S(T)$($ \equiv B(T)^{-1}$):
          $S(T)/S(0)$ versus $T/T_c$. Here the dotted line
          indicates the analytic formula (\protect\ref{actiontc})
          in the crossover temperature region, $|T-T_c| << T_c$.}
 \label{figtem}
\end{figure}

\begin{figure}
 \caption{Plots of $\ln \Gamma/\Gamma(0)$ as a function of 
          $T/T_c$, where $\Gamma(0)$ is the tunneling rate at
          $T=0$. (a) the numerical calculation in the
          entire temperature, (b) Eq.
          (\protect\ref{pthermal})
          for the pure thermal activation regime, (c) Eq.
          (\protect\ref{thermalrate1}) 
          for the quantum corrections, (d) Eq.
          (\protect\ref{thermaltc}) 
          for the quantum corrections close to the 
          crossover region, and (e) the rate in the crossover region 
          whose exponent is of the form
          (\protect\ref{actiontc}).}
 \label{figplot}
\end{figure}

\begin{figure}
 \caption{The $\thH$ dependence of the scaled crossover
          temperature $\bar{T}_c[=T_c(\thH)/T_c(135^{\circ})]$
          at (a) $k_2=0$, (b) 1.0, (c) 2.0, and (d) 100.}
 \label{fig:tcbar}
\end{figure}

\begin{figure}
 \caption{The dependence of $\theta_H^{T^{\rm max}_c}$ on $k_2$ which is
          derived from Eq. (\protect\ref{maxangle})
          where $k_2$ is defined in Eq.
          (\protect\ref{eq:dimpar}).}
 \label{fig:tcmax}
\end{figure}


\begin{references}
\bibitem[a]{kim_address}
Electronic address: gkim@phy.sejong.ac.kr

\bibitem{chu} 
M. Enz and R. Schilling, J. Phys. C
{\bf 19}, L711 (1986); J. L. van Hemmen and A. S\"{u}t\"{o}, Physica B
{\bf 141}, 37 (1986); G. Scharf, W. F. Wreszinski, and J. L. van Hemmen,
J. Phys. A {\bf 20}, 4309 (1987); E. M. Chudnovsky and L. Gunther, 
Phys. Rev. Lett. {\bf 60}, 661 (1988); P. Politi, A. Rettori,
F. Hartmann-Boutron, and J. Villain, Phys. Rev. Lett.
{\bf 75}, 537 (1995).

\bibitem{her} 
D. D. Awschalom, J. F. Smyth, G. Grinstein, D. P. Vincenzo,
and D. Loss, Phys. Rev. Lett. {\bf 68}, 3092 (1992);
L. Thomas, F. Lionti,
R. Ballou, D. Gatteschi, R. Sessoli and B. Barbara, Nature {\bf 383}, 145
(1996); J. R. Friedman, M. P. Sarachik, J. Tejada, and
R. Ziolo, Phys. Rev. Lett. {\bf 76}, 3830 (1996); 
J. M. Hernandez, X. X. Zhang, F. Luis, J. Tejada,
J. R. Friedman, P. Sarachik, and R. Ziolo, Phys. Rev. B {\bf 55},
5858 (1997).

\bibitem{gar} A. Garg, Phys. Rev. B {\bf 51}, 15592 (1995).

\bibitem{mig} M.-C. Miguel and E. M. Chudnovsky, Phys. Rev. B
{\bf 54}, 388 (1996).

\bibitem{kim} G.-H. Kim and D. S. Hwang, Phys. Rev. B
{\bf 55}, 8918 (1997).

\bibitem{zas} O. B. Zaslavskii, Phys. Rev. B {\bf 42}, 992 (1990).

\bibitem{lan} J. S. Langer, Ann. Phys.(NY) {\bf 41}, 108 (1967);
S. Coleman, Phys. Rev. D {\bf 15}, 2929 (1977).

\bibitem{del} J. von Delft and C. L. Henley,
Phys. Rev. Lett. {\bf 69}, 3236 (1992);
D. Loss, D. P. DiVincenzo, and G. Grinstein, {\it ibid.}
{\bf 69}, 3232 (1992); E. Fradkin and M. Stone,
Phys. Rev. B {\bf 38}, 7215 (1988). 

\bibitem{nie} H. B. Nielsen
and D. Rohrlich, Nucl. Phys. B {\bf 299}, 471 (1988);
K. Johnson, Ann. Phys. (N.Y.) {\bf 192}, 104 (1989);
J.-Q. Liang, H. J. W. Muller-Kirsten, J.-G. Zhou, F. Zimmerchied,
and F.-C. Pu, hep-th-9612106.

\bibitem{sto} E. C. Stoner and E. P. Wohlfarth, Philos. Trans.
Roy. Soc. London, {\bf A240}, 599 (1948).

\bibitem{comact} The effective moment of inertia is
derived from the partition function (\ref{amp}), 
the measure (\ref{mea}) and
the action (\ref{eq:newactn}) by performing the Gaussian integration 
over $\phi$.

\bibitem{comalt} Alternatively, if we consider {\bf M} as a spin 
operator, the Hamiltonian obtained from Eq. (\ref{eq:toten}) simply 
commutes with $M_z$ in the absence of $K_2$ and $H_x$, in which
QTM is not expected.\cite{mig}

\bibitem{comzero} Care should be taken in case of $k_2=0$
for $\thH=\pi$ because the terms
except for the first and last one in 
Eq. (\ref{eq:newactn}) vanish.
Since $\phi_{\rm cl}$ is not small in this situation,
we need to solve the coupled equation of motion for 
$\delbar_{\rm cl}$ and $\phi_{\rm cl}$ from Euler-Lagrange
equation. By using $\bar{E}(\delbar_{\rm cl},\phi_{\rm cl})=0$,
we get $\delbar_{\rm cl}=0$ and 2. 
$\phi_{\rm cl}$ can be any value for $\delbar_{\rm cl}=0$, 
but it does not 
contribute to the classical action. Since we obtain
$d\phi/d\tilde{\tau}=-i\epsn$ for $\delbar_{\rm cl}=2$,
$S^{\rm cl}_{E}$ becomes infinity. Thus, QTM is
not observable.\cite{kim}

\bibitem{chu1} E. M. Chudnovsky, Phys. Rev. A {\bf 46}, 8011 (1992);
D. A. Gorokhov and G. Blatter, Phys. Rev. B {\bf 56}, 3130 (1997).

\bibitem{ive} B. I. Ivelv, Yu. N. Ovchinnikov, and R. S.
Thompson, Phys. Rev. B {\bf 44}, 7023 (1991);
E. M. Chudnovsky, A. Ferrera
and A. Vilenkin, {\it ibid.} {\bf 51}, 1181 (1995).

\bibitem{aff} I. Affleck, Phys. Rev. Lett. {\bf 46}, 388 (1981);
D. Waxman and A. J. Leggett, Phys. Rev. B {\bf 32}, 4450 (1985);
W. M. Miller, J. Chem. Phys. {\bf 62}, 1899 (1975).

\bibitem{wei} U. Weiss, {\it Quantum Dissipative 
Systems}, (World Scientific, Singapore, 1993).

\bibitem{gra} H. Grabert, P. Olschowski, and U. Weiss, Phys. Rev.
B {\bf 36}, 1931 (1987). 

\bibitem{comcross} In Fig. \ref{figplot}, we make use of the tunneling
rate in the crossover region which is of the form
\begin{eqnarray}
\Gamma={\omega_b \over 2 \pi} A_0 
{\sinh(\beta \hbar \omega_0/2) \over \pi} \sqrt{\pi} \kappa ~
{\rm erfc}(-\kappa \varepsilon) \exp [ (\kappa \varepsilon)^2
-\beta U_m],
\nonumber
\end{eqnarray}
where\cite{wei} $\kappa^2={18 \over 5} \beta U_m$, 
$\varepsilon \equiv 1-T/T_c$, 
${\rm erfc}(x)={2 \over \sqrt{\pi}} \int^{\infty}_{x} dt \exp(-t^2)$,
and $A_0 =1 $ for $T \geq T_c$ and
$T/T_c$ for $T \leq T_c$.

\bibitem{leg} A. J. Leggett, in {\it Quantum Tunneling
of Magnetization-QTM '94}, edited by L. Gunther and
B. Barbara (Kluwer Academic, Dordrecht/Boston/London, 1995).

\bibitem{wer1} W. Wernsdorfer, E. Bonet Orozco, K. Hasselbach,         
A. Benoit, D. Mailly, O. Kubo, H. Nakano and B. Barbara, Phys. Rev.
Lett. {\bf 79}, 4014 (1997).

\bibitem{chu2} E. M. Chudnovsky, in {\it Quantum Tunneling
of Magnetization-QTM '94}, edited by L. Gunther and
B. Barbara (Kluwer Academic, Dordrecht/Boston/London, 1995).

\bibitem{boz} R. M. Bozorth, in {\it Ferromagnetism} (IEEE,
New York, 1978).

\bibitem{bar} B. Barbara, W. Wernsdorfer, L. C. Sampaio, J. G. Park,
C. Paulsen, M. A. Novak, R. Ferr\'e, D. Mailly, R. Sessoli, 
A. Caneschi, K. Hasselbach, A. Benoit, L. Thomas, 
J. Magn. Magn. Mater. {\bf 140-144}, 1825 (1995); H. Yamazaki, G. Tatara,
K. Katsumata, K. Ishibashi, Y. Aoyagi, {\it ibid.} 
{\bf 156}, 135 (1996);
X. X. Zhang, J. M. Hernandez, J. Tejada, and R. F. Ziolo,
Phys. Rev. B {\bf 54}, 4101 (1996).   

\end{references}
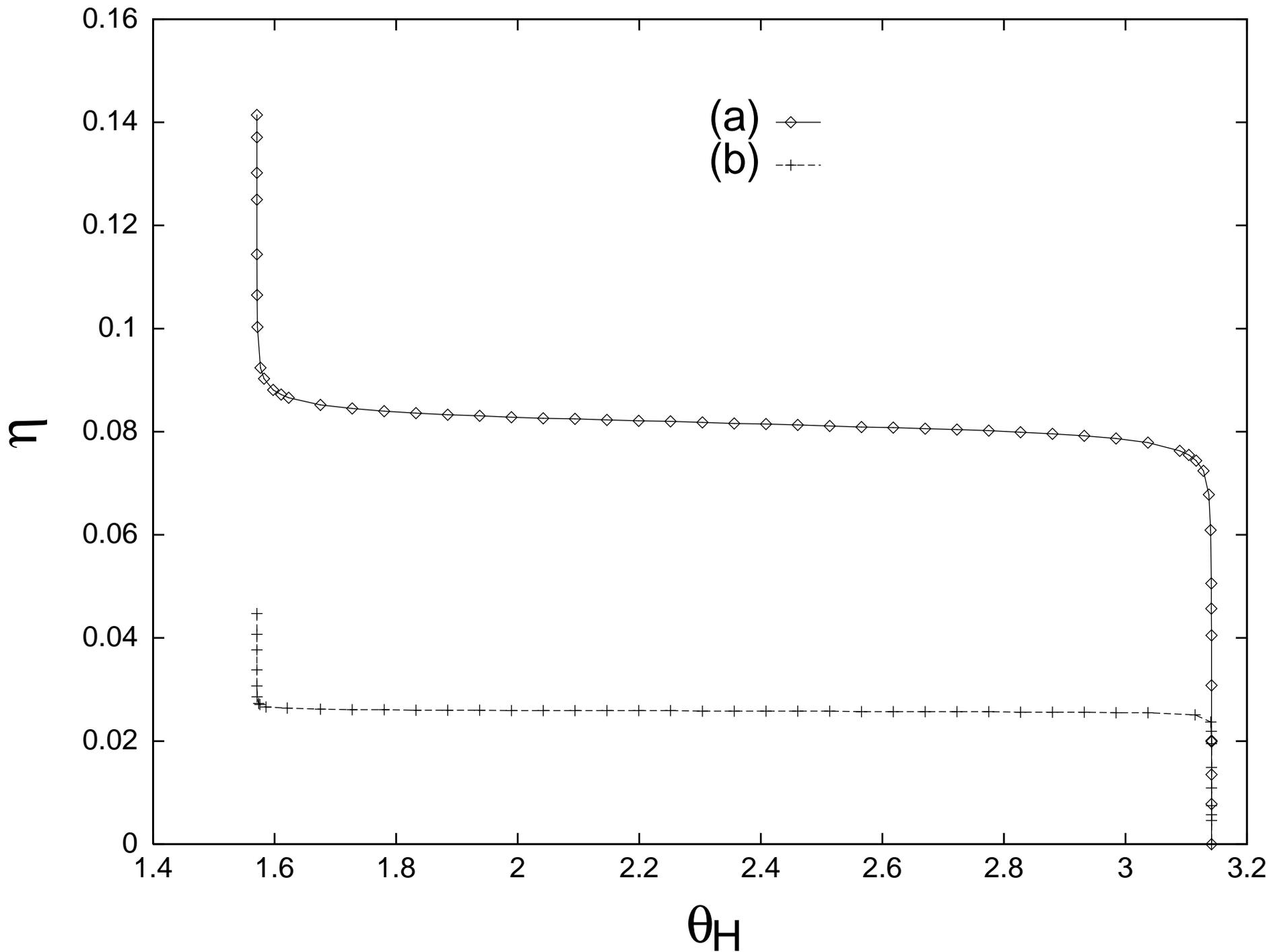
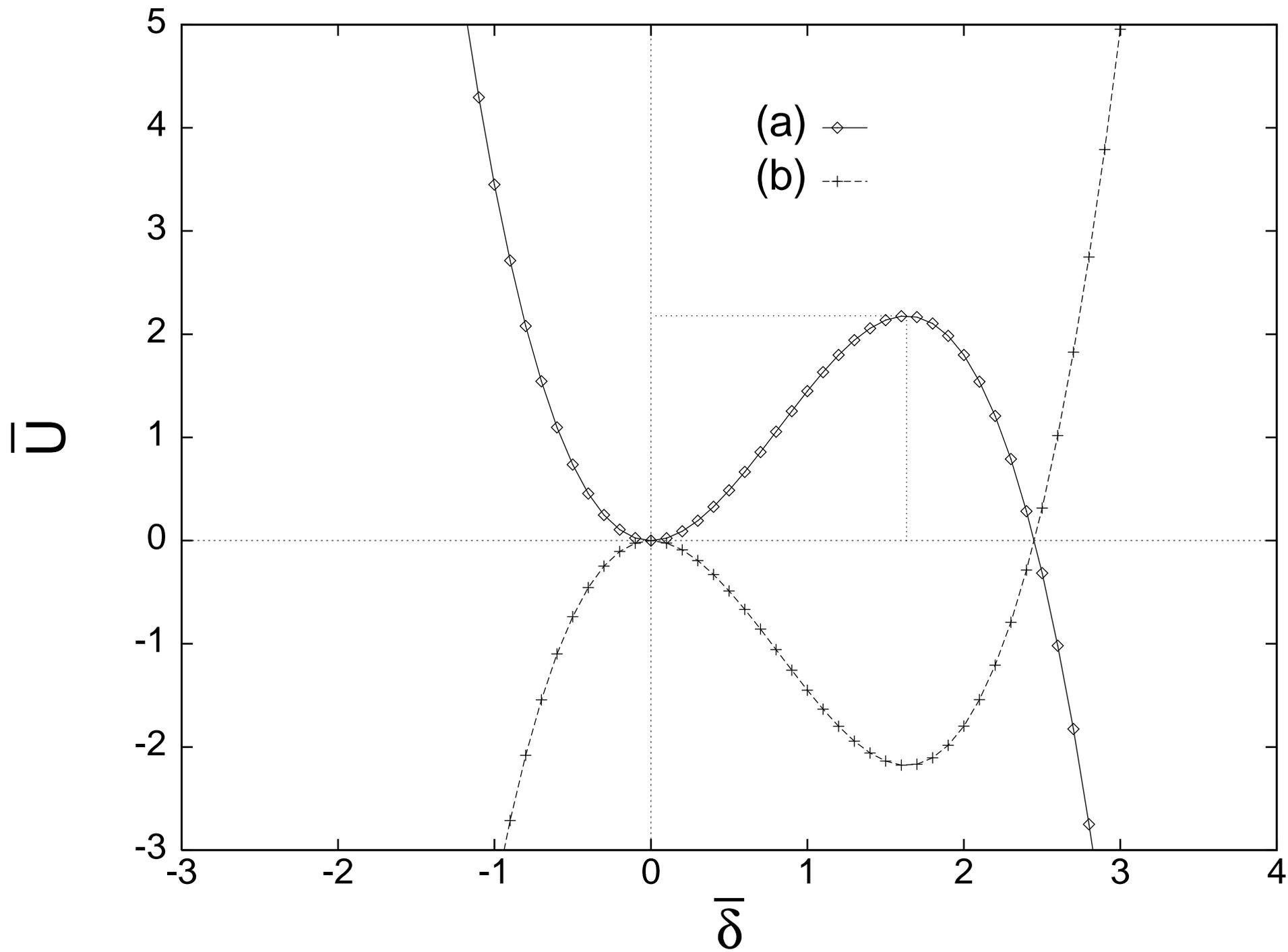
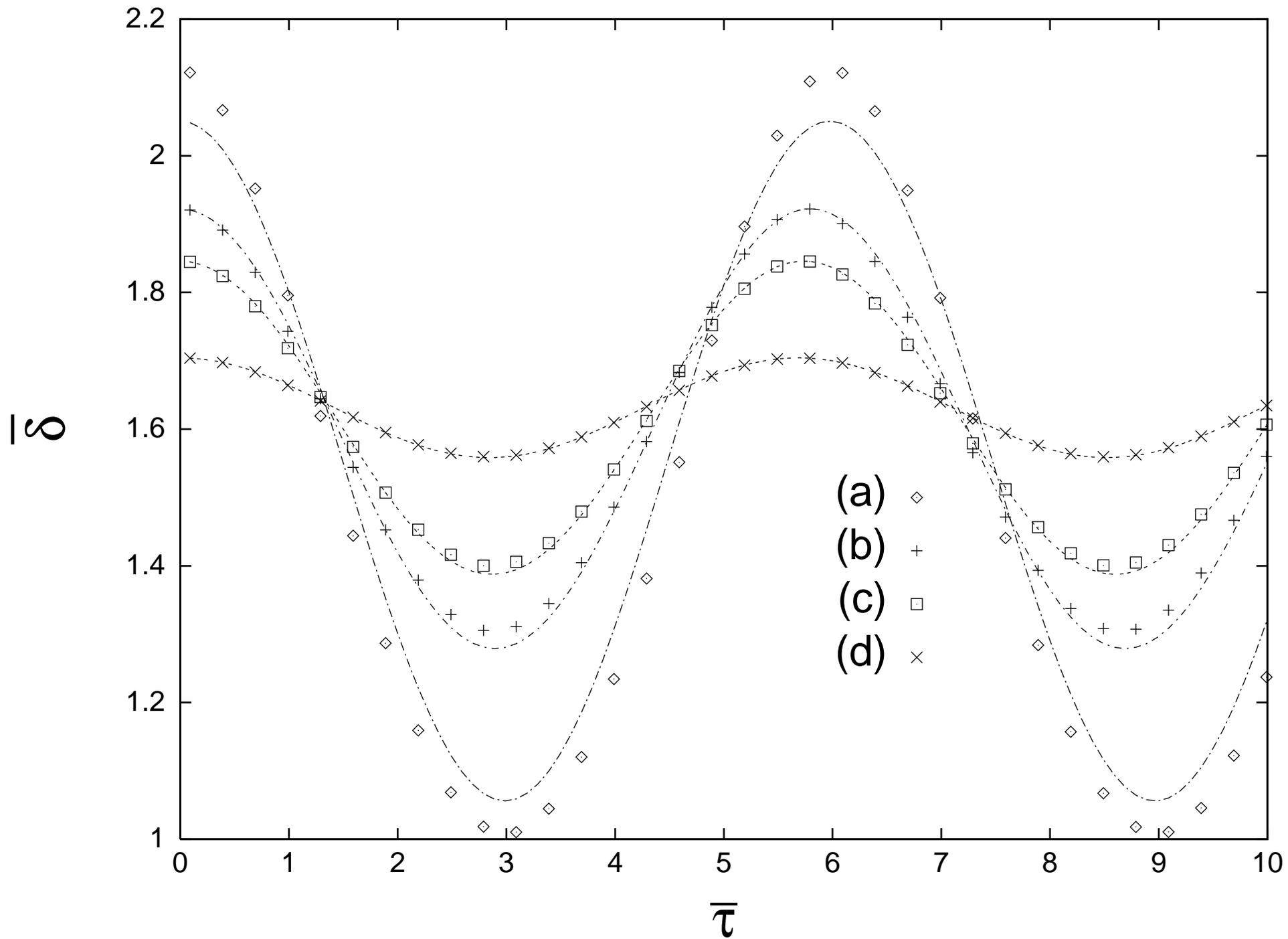
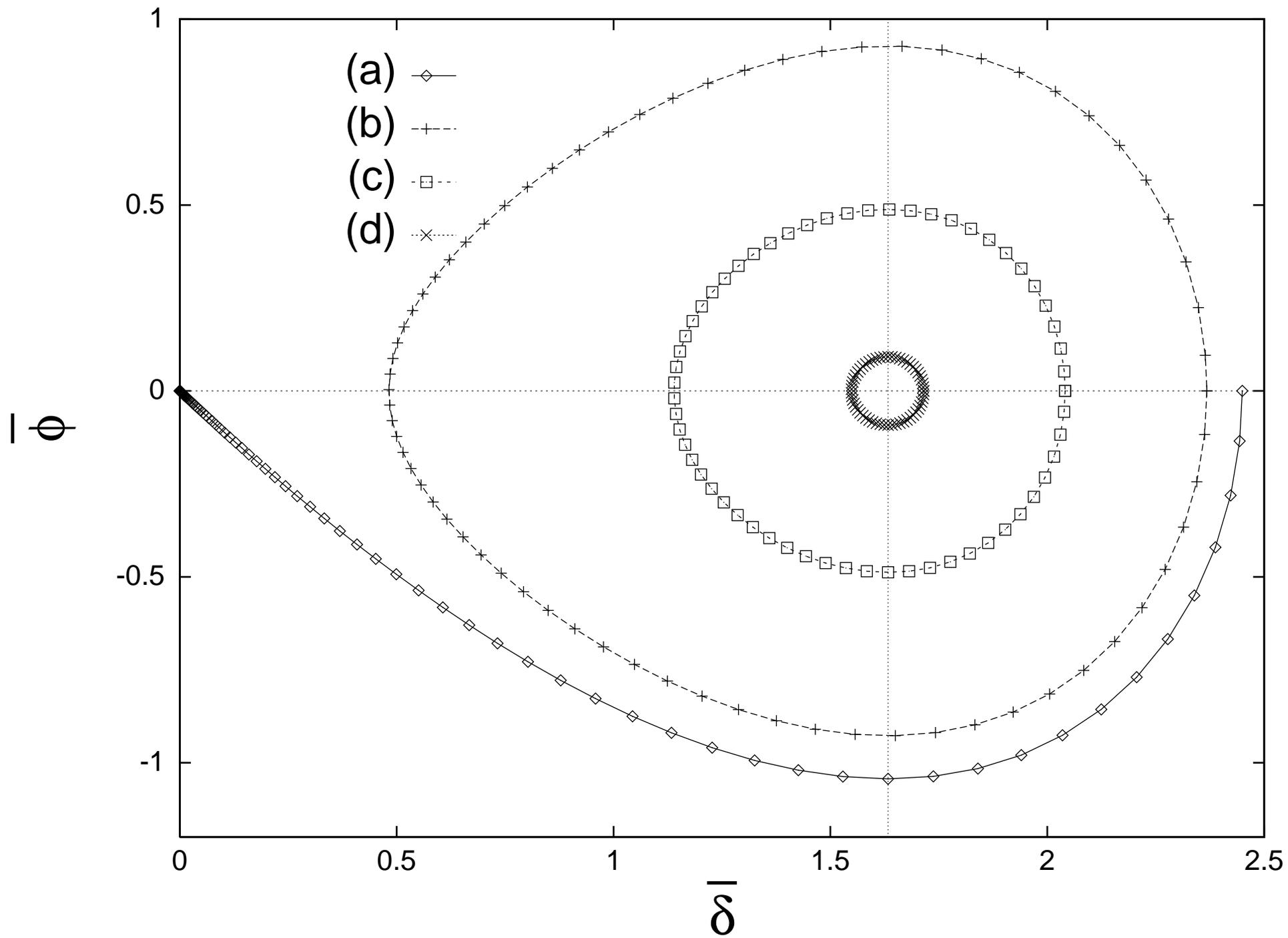
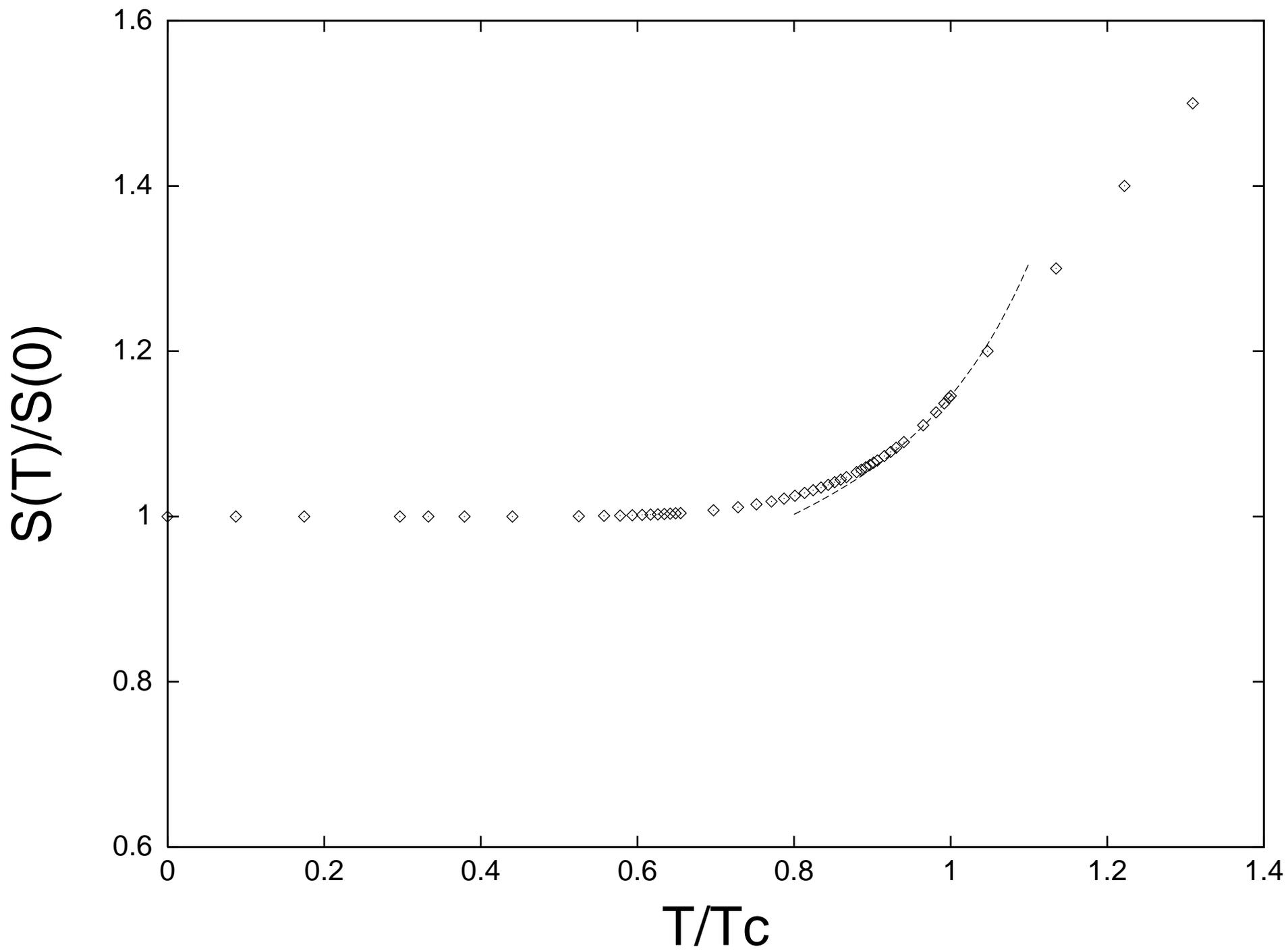
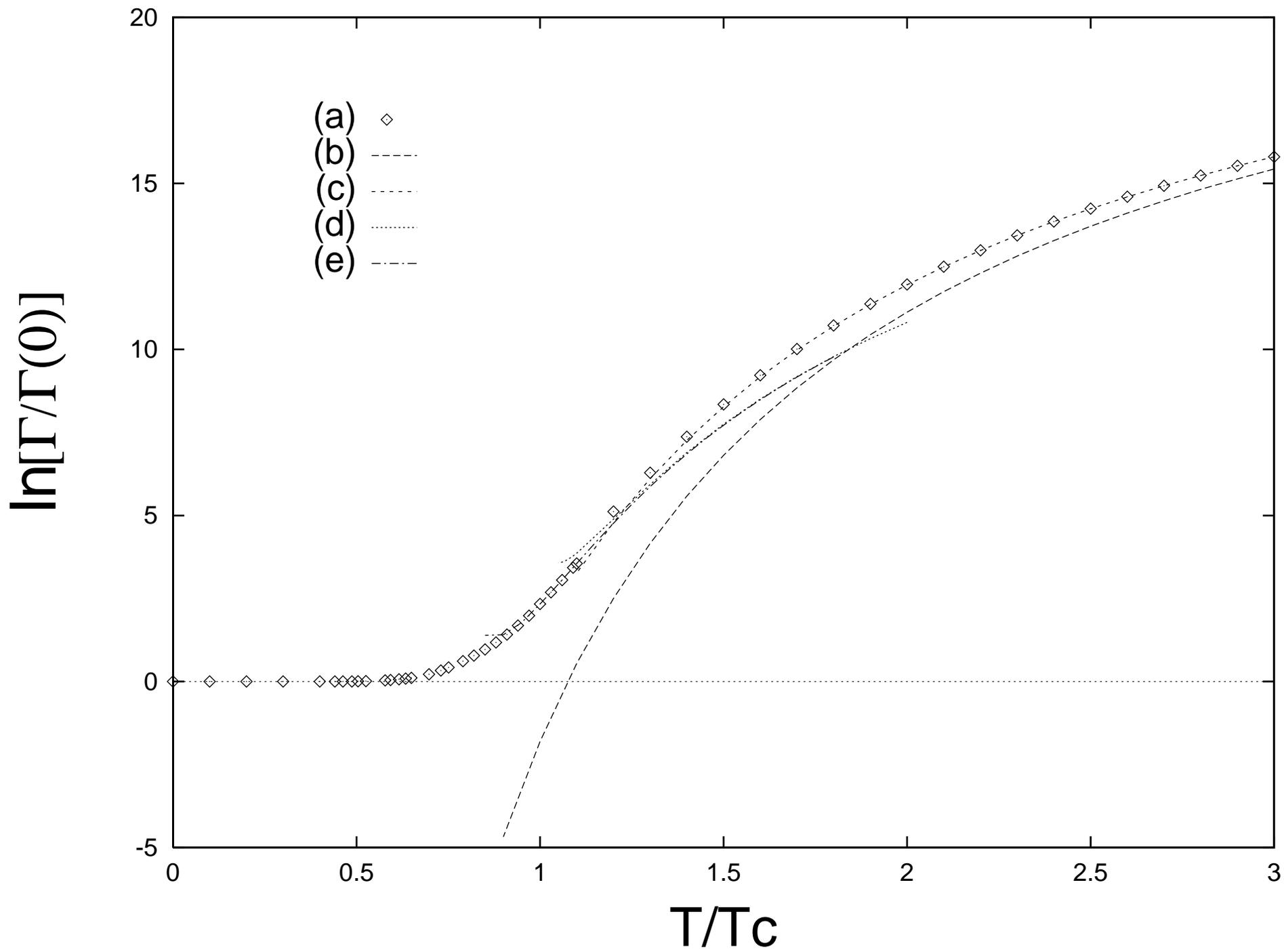
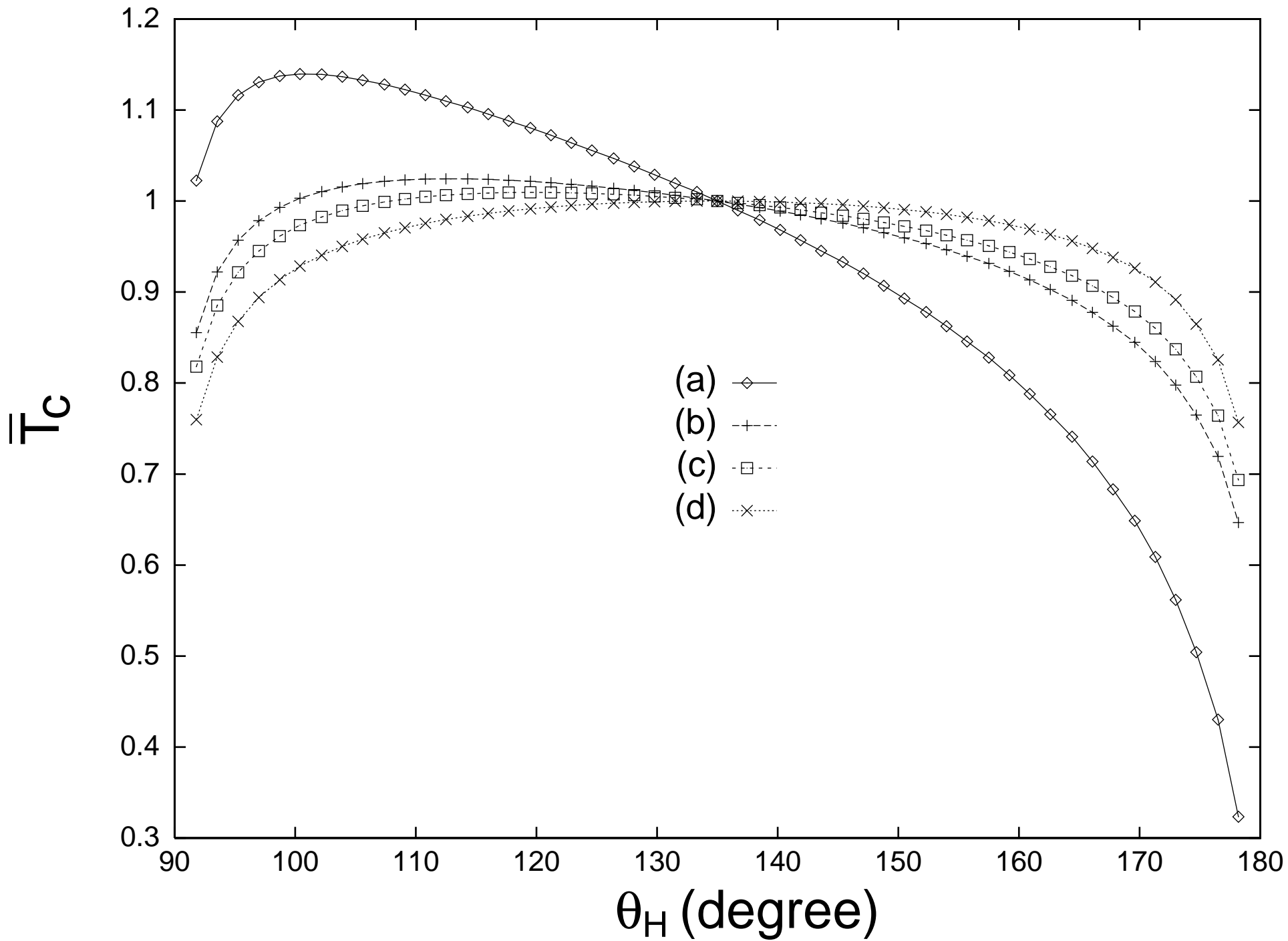
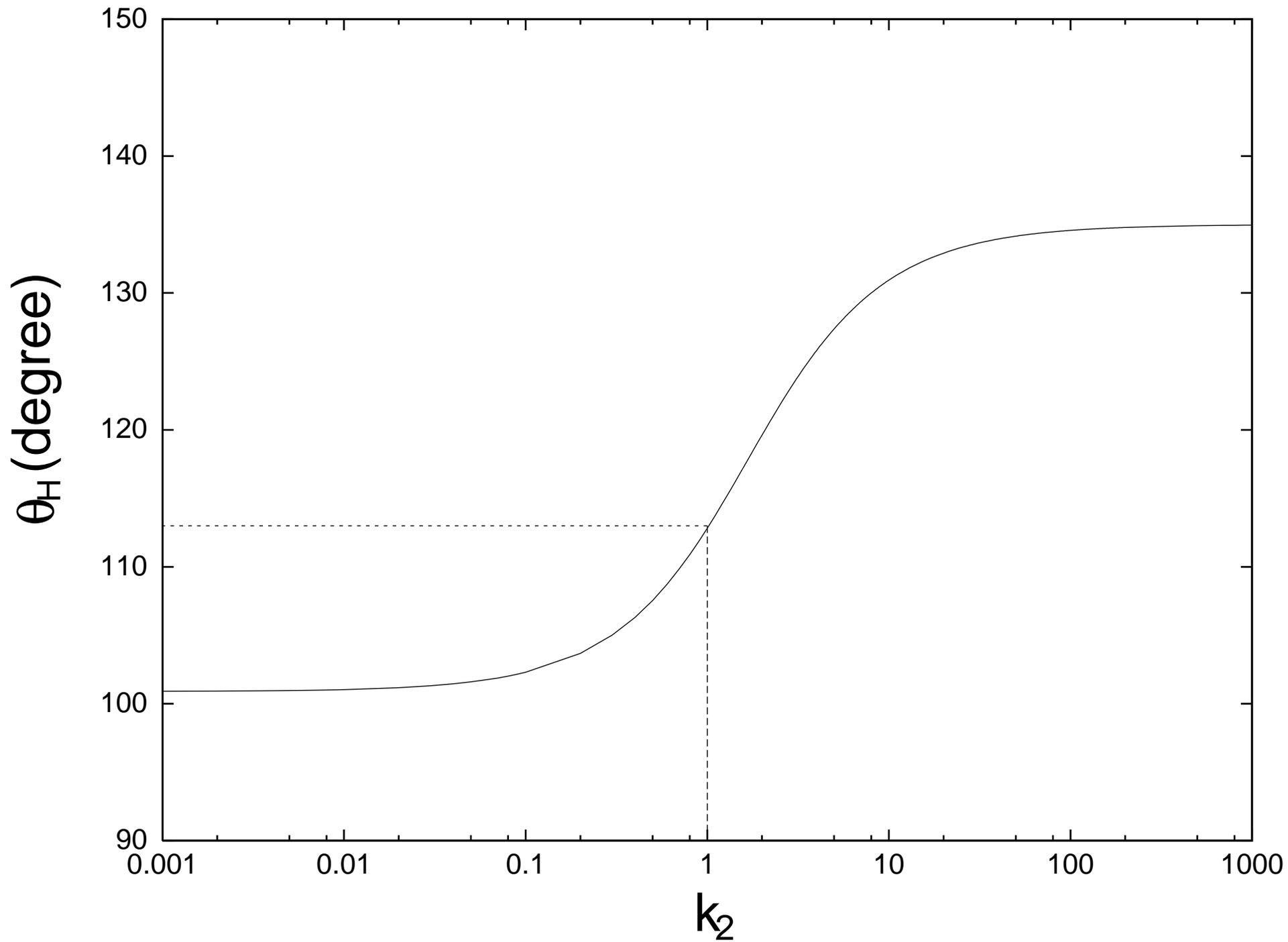
\end{document}